# Supporting quantum technologies with an ultra-low-loss silicon photonics platform

Matteo Cherchi[*], Arijit Bera, Antti Kemppinen, Jaani Nissilä, Kirsi Tappura, Marco Caputo, Lauri Lehtimäki, Janne Lehtinen, Joonas Govenius, Tomi Hassinen, Mika Prunnila, and Timo Aalto

VTT – Technical Research Centre of Finland Ltd, Tietotie 3, 02150 Espoo, Finland

**ABSTRACT**

Photonic integrated circuits (PICs) are expected to play a significant role in the ongoing second quantum revolution, thanks to their stability and scalability. Still, major upgrades are needed for available PIC platforms to meet the demanding requirements of quantum devices. In this paper, we present a review of our recent progress in upgrading an unconventional silicon photonics platform towards this goal, including ultra-low propagation losses, low fibre coupling losses, integration of superconducting elements, Faraday rotators, fast and efficient detectors, and phase modulators with low loss and/or low energy consumption. We show the relevance of our developments and of our vision in two main applications: quantum key distribution – to achieve significantly higher key rates and large-scale deployment – and cryogenic quantum computers – to replace electrical connections to the cryostat with optical fibres.

## 1. INTRODUCTION

We are presently living the so-called second quantum revolution, where the focus has shifted from pure science to technologies and applications[1]. Photonic technologies are expected to play a major role, not only in quantum applications, but also in classical configurations to support solid-state quantum systems. In particular, photonic integrated circuits (PICs) offer unique opportunities for different quantum technologies to scale up system complexity and integration density while providing unmatched performance and stability[2–6]. In this regard, the micron-scale silicon photonics platform[7] brings with it a unique set of properties and building blocks. This includes low propagation losses (down to 3 dB/m demonstrated to date[8,9]), broadband and low-loss coupling to fibres (≈ 0.5 dB), fast (> 40 GHz) and responsive (≈ 1 A/W) integrated germanium photodetectors[10], up-reflecting mirrors for broadband and low-loss coupling to arrays of detectors, tight bends[11] enabling high integration density, efficient phase shifters, low-loss Mach-Zehnder interferometers, multi-million Q ring resonators[9,12], polarization insensitive operation, and polarization splitters[13] and rotators, including all-silicon Faraday rotators[14].

A relevant example application is large-scale deployment of quantum key distribution (QKD), for which we are developing efficient multiplexed receivers. A second interesting case is the use of our photonic integration technology to scale up superconducting quantum computers by controlling and reading out the qubits in the cryostat through classical optical links. In this case the major challenge is the development of suitable electrical-to-optical and optical-to-electrical converters operating at cryogenic temperatures.

In the following, we will cover these ongoing developments showing our recent results as well as our plans to further exploit the platform. In Section 2, we will first give an overview of the thick-silicon photonics platform, with a special focus on the most relevant features for quantum technologies. In Sections 3 and 4, we will cover the ongoing developments for QKD and quantum computers and then conclude in Section 5 and briefly mention other promising future developments and applications.

## 2. OVERVIEW OF VTT THICK SOI PLATFORM

We can divide the building blocks of the platform into two main categories: passives and actives. In this context, "active" means anything requiring an electrical control – like thermo-optic phase shifters and electro-optic modulators – or electrical readout – like a photodiode. An overview of the main building blocks available on the platform is sketched in Figure 1 with the notable exception of phase modulators based on PIN diodes, which are explained in detail in §2.2. We fabricate our PICs on 150 mm diameter silicon-on-insulator (SOI) wafers (to be upgraded soon to 200 mm) with 3 µm thick device layer (± 100 nm uniformity) using a UV stepper (365 nm wavelength) and a modified Bosch process[15,16] to etch the waveguides.

*matteo.cherchi@vtt.fi; phone +358 40 684 9040, ORCID 0000-0002-6233-4466, www.vtt.fi

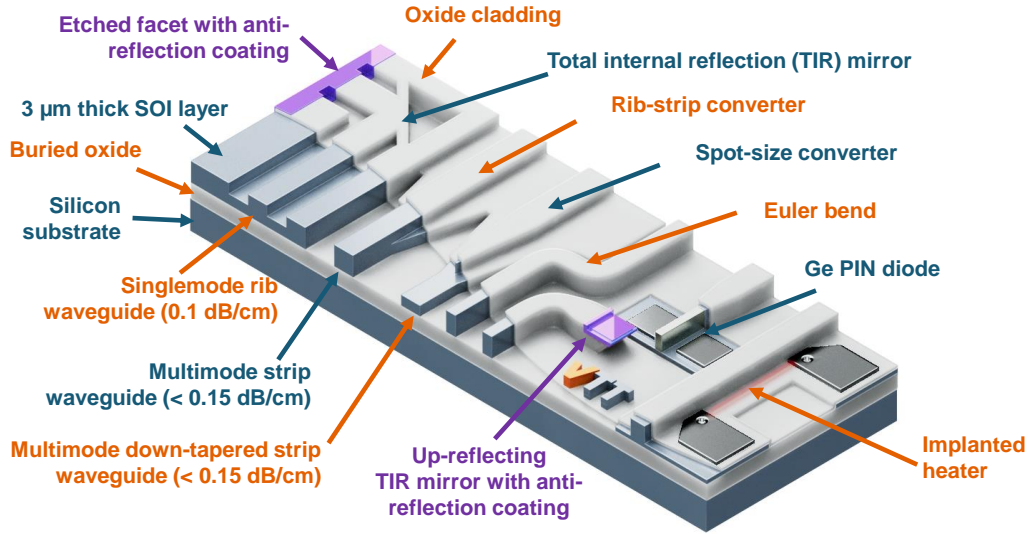

Figure 1. Sketch of the main building blocks available on the thick-SOI platform. Typical thickness of the device layer is 3 µm, whereas the buried oxide (BOX) thickness can vary from 400 nm to 3 µm. We define "active" building blocks as those requiring electrical pads for either control or readout.

## 2.1 Passive building blocks

*Types of waveguides*

The five main waveguide types available on the platform are: rib waveguides, strip waveguides, down-tapered strip waveguides, strip waveguides with a thin pedestal, and down-tapered strip waveguides with a thin pedestal (the latter is the only type missing in Figure 1). For rib waveguides, trenches are partially etched (typically 1.2 µm deep etch) on both sides of the waveguide. Single-mode operation can be achieved for both transverse-electric (TE) and transverse-magnetic (TM) polarisation with a suitable choice of the rib width[17] (typically ≤ 3 µm). On the contrary, all four possible strip waveguide cross sections are inherently multimode. Nevertheless, we carefully design the optical circuits so that excitation of the higher order modes (HOMs) is always negligible in the connecting waveguides, ensuring effective single-mode operation of the whole circuit.

*I/O coupling*

We fabricate the vertical waveguide facets of our PICs at wafer scale by first etching the silicon facet and then depositing a suitable anti-reflection coating, which can be made of either a single dielectric layer or multiple layers.

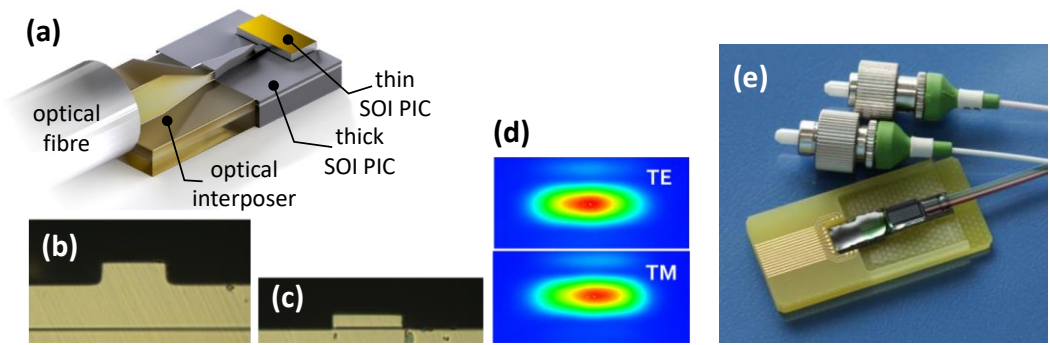

Figure 2. (a) Sketch of different mode size conversion starting from a SMF coupled to the 3 µm thick waveguides of a thick SOI PIC using an optical interposer fabricated on 12 µm thick SOI. The sketch shows also how the mode size can be reduced further even to couple light to submicron waveguides on a flip-chip bonded PIC that can be evanescently coupled through suitable inverse tapers; (b) micrograph of 12 µm thick rib waveguide of a fabricated optical interposer and (c) micrograph of a strip waveguide polished down to about 3 µm thickness on the opposite facet; (d) near field image (infrared camera) of the TE and TM modes at the output facet of the interposer (shown in (c)); (e) packaged 3 µm thick SOI PIC coupled to a fibre array through an optical interposer.

The coupling loss to optical fibres can be as low as 0.5 dB, provided that the mode field diameter is about 2.5 µm, which is achieved with lensed (or tapered) fibres or small core fibres with high numerical aperture. Fibre arrays must be used instead in configurations where the PIC has several inputs and outputs. Given the limited assembly precision of fibre arrays and considering that the tolerance to misalignments scales inversely with the mode size, low loss coupling can be ensured only by arrays of standard single mode fibres (SMFs) with mode diameter around 10 µm. This requires suitable mode size converters, like the one shown in Figure 2, fabricated by etching arrays of 12 µm wide rib and strip waveguides on a SOI wafer with a 12 µm thick device layer, and then tapering the thickness of the output strip waveguides down to 3 µm by polishing each optical interposer chip. We are presently working towards further reduction of coupling losses below 0.5 dB by implementing 3D printed lenses[18], directly printed on the waveguide facets at wafer scale. This way, we aim to couple the PIC directly to SMFs and fibre arrays with ultra-low loss and relaxed alignment tolerance[19].

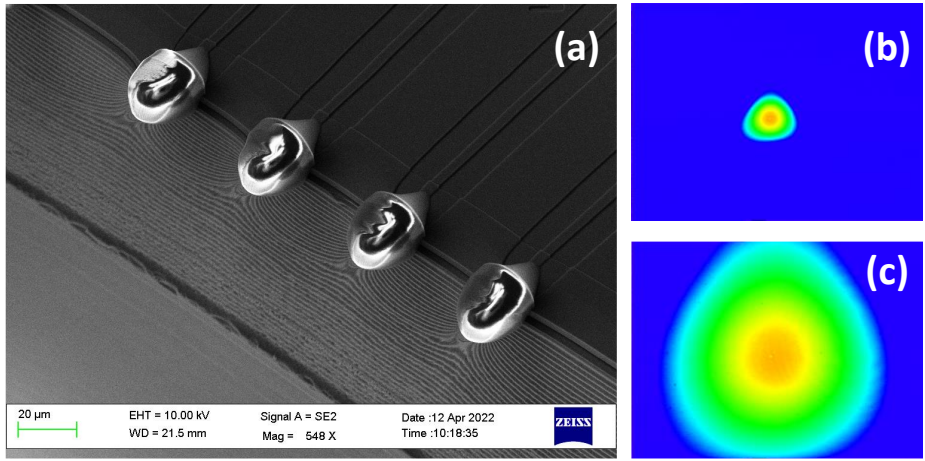

Figure 3. a) SEM image of polymer lenses 3D printed in front of the end facets of four rib waveguides; b) near-field picture of the output mode of a rib waveguide taken with an infrared camera; c) near-field picture of the output of a lensed rib waveguide (same scale as b)).

Light can also be coupled to the PIC vertically from up-reflecting mirrors (URMs, see Figure 1 and Figure 4) which are wet etched with a negative angle. Their working principle is total internal reflection, and coupling losses are practically the same as for the vertical facets. The anti-reflection coating for URMs is the same as for vertical facets. The wet etching process occurs along crystalline planes, meaning that the mirrors can be fabricated only along the four orthogonal crystal planes with Miller indices $110, \bar{1}10, 1\bar{1}0$, and $\bar{1}\bar{1}0$. One of the advantages of URMs is the possibility to use them to test the fabricated PICs at wafer scale.

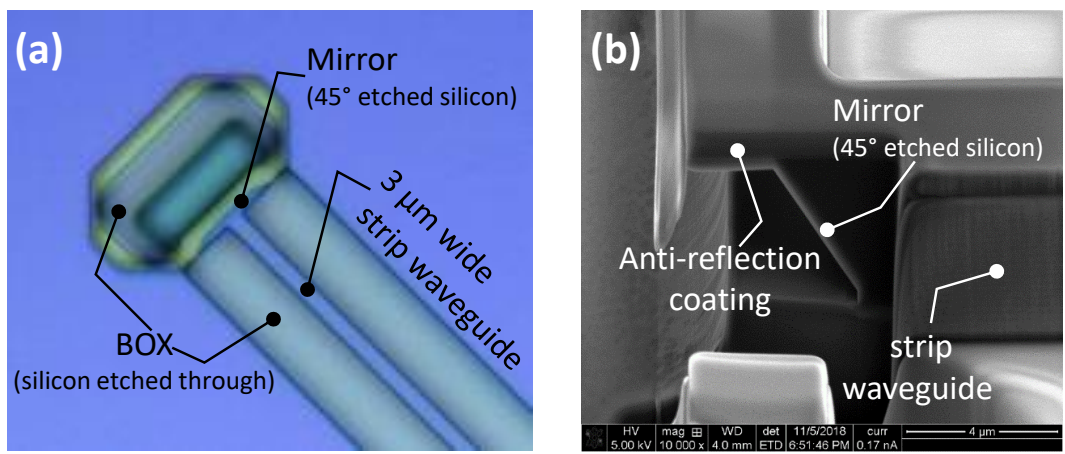

Figure 4. (a) Micrograph of a fabricated URM and (b) side view of a vertical cross-section of an URM via focused ion beam microscopy.

Compared to grating couplers typically used in submicron waveguides, URMs support both TE and TM polarisations with negligible polarisation-dependent loss, and they operate over the whole transparency range of silicon, from 1.2 µm to 7 µm wavelength. We additionally stress here that thick SOI PICs can operate over the whole transparency range of silicon. In particular, one can design a rib waveguide to be single-mode at all the wavelengths in that broad spectral range, spanning several octaves[17]. We must emphasize that, above 3 µm wavelength, absorption in the silica cladding (including the buried oxide and the top cladding, see Figure 1) contributes to propagation losses. However, due to the strong confinement in the thick silicon core, propagation losses remain below 1 dB/cm up to about 4 µm wavelength, and low propagation losses can be achieved up to 7 µm wavelength by selectively removing the silica cladding[20] all around the waveguide. This is in strong contrast to thin SOI-based Si photonics platforms where propagation losses are typically above 1 dB/cm even at telecom wavelengths[20] because of stronger interaction with the sidewall roughness. Furthermore, the power fraction in the cladding of submicron waveguide modes is orders of magnitude larger compared to thick SOI waveguides, meaning that the absorption in the silica cladding[21] becomes unbearable already beyond 2.5 µm wavelength. Operation in the mid-IR is critical for many gas sensing applications[22] – including quantum sensing[23] – and also to exploit the strong third order nonlinearity of silicon, further, the larger optical mode size in thick SOI enables avoiding saturation caused by strong two-photon absorption[24] (and associated free-carrier absorption[25]) at wavelengths shorter than 2.2 µm. While the large cross section of the waveguides is not ideal for efficient excitation of nonlinear effects – including parasitic two-photon absorption – the unique combination with ultra-low propagation losses is advantageous in many applications[26,27].

*Tight bends enabling high integration density*

It is generally assumed that waveguides with micron scale cross-sections require bending radii in the order of several millimetres. This is because the index contrast ensuring single-mode operation in a micron scale waveguide would inherently lead to high radiation losses for tighter bends. In the platform, we have developed two solutions to this limitation: turning mirrors based on TIR[28] (Figure 5(a) and (b)) and tight adiabatic bends referred to as the Euler bends[11] (Figure 5(a)). The first approach applies to both rib waveguides and strip waveguides, whereas the second requires high index contrast strip waveguides.

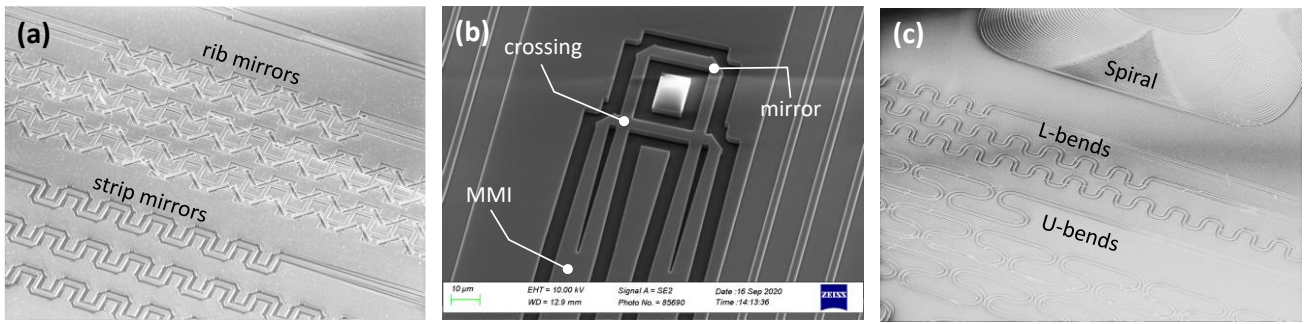

Figure 5. (a) SEM picture of 90° turning mirrors on rib waveguides and strip waveguides; (b) detail of a compact imbalanced MZI based on TIR mirrors; (c) SEM picture of Euler bends with L and U shape and detail of a spiral waveguide using larger L-bends.

TIR mirrors allow for compact layouts like the imbalanced Mach Zehnder interferometer (MZI) shown in Figure 5(b), which also shows the very low-loss (≈ 0.02 dB) waveguide crossings easily achievable on the platform. Turning mirrors can be designed with almost any turning angle, and their losses can be made as low as 0.1 dB per turn by using parabolic shapes and/or making the waveguide sufficiently wide. In fact, the main loss mechanism is light diffraction due to the partial lack of lateral guidance in the mirror region. Remarkably, turning mirrors work over the entire (1.2 to 7 µm) wavelength transparency region of silicon and can be designed to simultaneously work equally well for both TE and TM polarisation, despite the polarisation-dependent offset induced by the Goos-Hänchen shift[29]. Nevertheless, the non-negligible loss makes them unsuitable for circuits requiring a large number of bends, e.g., long spiral waveguides.

For this reason, we have also developed more conventional waveguide bends to achieve much lower losses. They are based on strip waveguides, ensuring negligible radiation loss due to strong light confinement. The only limitation is that they support several HOMs that get easily excited in a tight bend. We have therefore introduced[11] and patented[30] a geometry with gradual a change of curvature, using the Euler spiral geometry as shown Figure 5(c) and Figure 6(a) and (b). This way, tight bends with loss lower than 0.02 dB can be achieved with effective bending radii of a few tens of microns, enabling, for example, compact race track resonators with quality factor Q of up to 14 million[9,12].

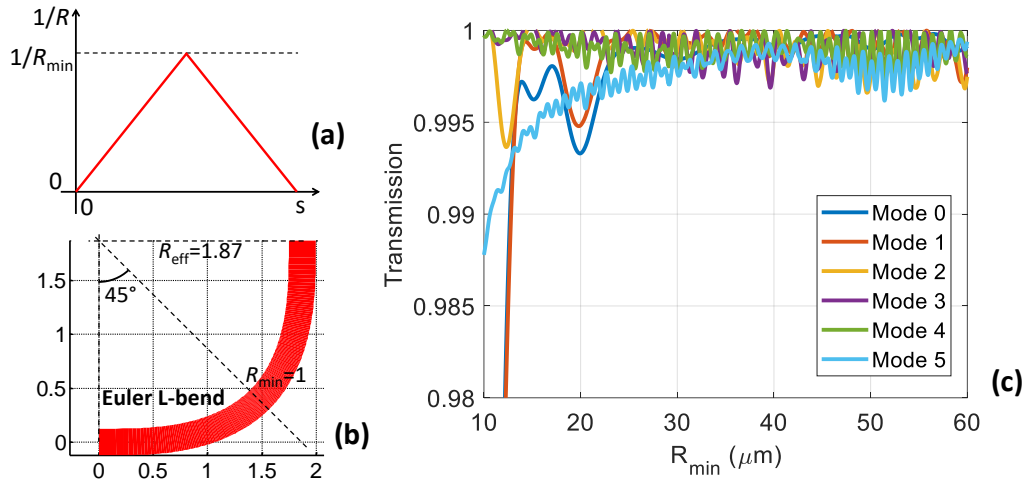

Figure 6. (a) Plot showing the linear change of the curvature $1/R$ as a function of the length $s$ in a Euler bend, starting from zero, reaching up to $1/R_{min}$ and then going back to zero symmetrically; (b) Example layout of a 90° Euler bend (or L-bend) with unity minimum bending radius, showing the resulting effective radius $R_{eff}$; (c) Simulation of the transmission of the $TE_{00}$ mode and of five horizontal higher order TE modes of a 1.5 µm wide strip waveguide at the output of a 90° Euler bend as a function of the minimum bending radius. The five HOM $TE_{n0}$ modes (n = 1…5) have $n$ nodes in the horizontal direction and zero nodes in the vertical direction. The wavelength is 1.55 µm.

Even though, in general, the wavelength range of operation of the bends is not as wide as that of turning mirrors, bends can be designed to cover bandwidths of several hundreds of nanometres up to a few microns. An interesting property of Euler bends is that they very efficiently transmit most of supported HOMs[31] (i.e., those with effective index sufficiently higher than the cladding refractive index), as highlighted in Figure 6(c). In other words, the bends preserve the mode power distribution, which is useful when designing PICs for mode multiplexing[32] and when using spatial modes as a quantum degree of freedom[33–36] (see also §5). It is worth mentioning that turning mirrors also preserve the HOM power distribution under reflection[37].

*Polarisation management*

Micron-scale silicon waveguides support both TE and TM polarisation with very similar spatial mode distribution, the same propagation losses, and very similar effective indices. Indeed, strip waveguides with a square cross-section can support TE and TM fundamental modes with identical propagation constants. Any possible residual birefringence induced by material strain can be easily compensated by fine-tuning the waveguide width.

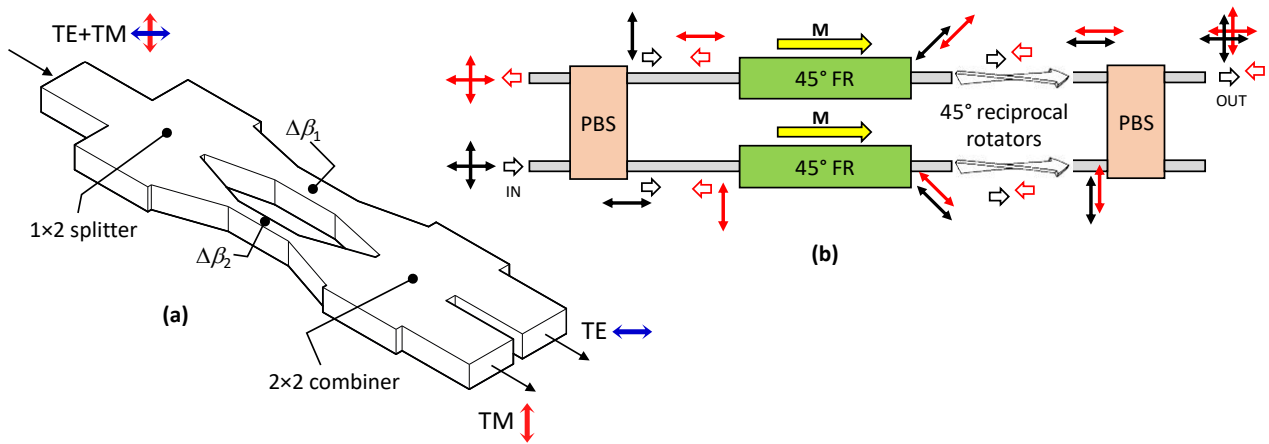

Figure 7. (a) Sketch of a MZI exploiting the form birefringence of waveguides of different width to serve as a PBS; (b) Scheme of a possible implementation of an integrated light circulator by combining PBSs, Faraday rotators (FR) and reciprocal polarization rotators on chip.

Most of the building blocks, including multimode interference (MMI) splitters, can be designed to support both polarisations at the same time. On the other hand, in many applications (including telecom and sensing) polarisation can be used as a degree of freedom, in which case a polarisation splitter/combiner is needed and preferably also different types of polarization rotators. We are presently developing a wide portfolio of building blocks for polarisation management, including MZI polarisation beam splitters (PBSs)[13,38] (see Figure 7(a)) and rotators[39]. Remarkably, we have demonstrated the use of silicon itself as magneto-optic material and achieved Faraday rotation in zero-birefringence waveguides[14]. Our ultimate goal is to build a fully-integrated all-silicon circulator based on splitters/combiners and reciprocal and non-reciprocal rotators (Figure 7(b)).

We conclude this section by mentioning that Faraday mirrors are commonly used in quantum photonics, including quantum key distribution systems (see §3), to ensure stable operation[40,41], as the polarization of the reflected light is always orthogonal to the input polarization[42] (i.e., antipodal on the Poincaré sphere). Faraday mirrors can be achieved in the platform by combining a Faraday rotator with a back-reflector like a MMI reflector or a Sagnac loop[43].

*Low-loss wavelength filters*

We have demonstrated several different types of wavelength filters, including ring resonators with Q up to 14 million[9,12], compact MMI resonators[43], flat-top lattice filters[44], and flat-top ring-loaded MZIs[45]. Some of these filters can be designed to have less than 0.5 dB excess loss. We have also demonstrated low-loss echelle gratings and arrayed waveguide gratings (AWGs)[46]. In Figure 8(a) we show the layout of an AWG with a small footprint due to the use of Euler bends. The device is polarisation independent because of the incorporation of strip waveguides with a square cross-section. Excess loss is in the 2 dB to 3 dB range, and the extinction ratio (ER) is larger than 25 dB. We have also demonstrated AWGs with loss in the 1 dB to 2 dB range[46], and ER exceeding 30 dB on all channels. We are presently working on further reduction of excess loss by improving design and fabrication of the star coupler. For echelle gratings like the one shown in Figure 8(a) we have already demonstrated excess loss lower than 1 dB for both polarisations with extinction ratio exceeding 20 dB for all channels.

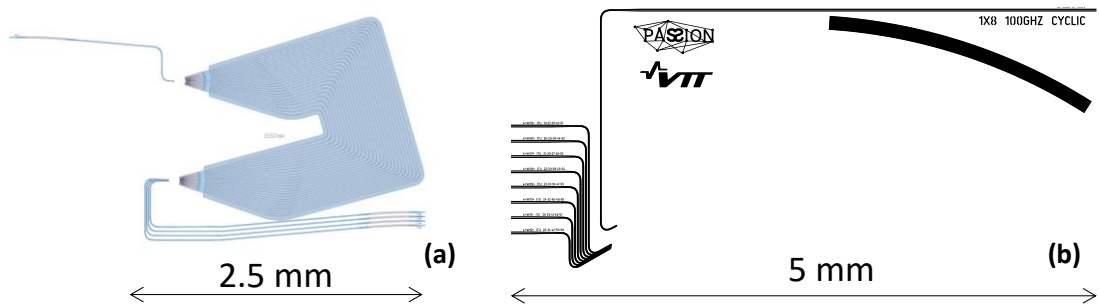

Figure 8. (a) Compact AWG with 100 GHz channel spacing and 5 nm free spectral range exploiting Euler bends and nearly-zero birefringence waveguides, ensuring polarisation independent operation; (b) cyclic echelle grating with 100 GHz channel spacing.

*Interfacing micro- and nano-scale devices*

The low propagation loss of micron-scale waveguide technology comes with the price of weak interaction with any element integrated directly on top of the waveguides. This problem has been successfully addressed with different fabrication techniques that do not jeopardize the performance of the platform. For example, recently we have been developing a light escalator[47] made of hydrogenated amorphous silicon (a-Si:H) to interface micron-scale waveguides with submicron waveguides, thin layers including 2D materials, and superconducting nanowires. We grow a submicron layer of a-Si:H (refractive index around 3.65) on top of the crystalline device layer (refractive index around 3.48 at 1550 nm wavelength), and pattern it to achieve adiabatic light coupling. The simulation in Figure 9(a) shows how the light propagates from the thick silicon waveguide to the thinner a-Si:H layer. The a-Si:H layer thickness can be optimised to maximise the overlap of the propagating light with, e.g., a graphene layer (for applications such as light detection[48] or modulation[49]) or a superconducting nanowire single photon detector[50] (SNSPD, see also §2.2) sandwiched between crystalline silicon (c-Si) and a-Si:H, similar to what is sketched in Figure 9(b). Furthermore, crystalline silicon can also be selectively removed and replaced with a deposited silica layer before depositing the a-Si:H layer, as sketched in Figure 9(c). The resulting high index contrast a-Si:H waveguide allows us to interface the micron-scale waveguide with submicron waveguides including plasmonic slot waveguides or even just PICs based on submicron silicon waveguides that can be simply bonded on top of the a-Si:H waveguide and evanescently coupled via inverse tapers. Both types of escalators can be fabricated using the same fabrication process. Another unique opportunity to couple microscale waveguides to nanophotonic devices comes from the URM. In fact, there are cases requiring the light to propagate across a functional surface (unlike the escalator case, where it propagates along it). In these cases, functional surfaces can be fabricated or just transferred on top of the flat output surface of the mirror (which is made of crystalline smooth silicon, not etched). This is a straightforward way to integrate metasurfaces including waveplates[51], metalenses[52], or electro-optic modulators[53,54]. However, the limited size of the mirror (3 µm in the waveguide propagation direction, i.e., a few wavelengths) may present a challenge for the design of the metasurface.

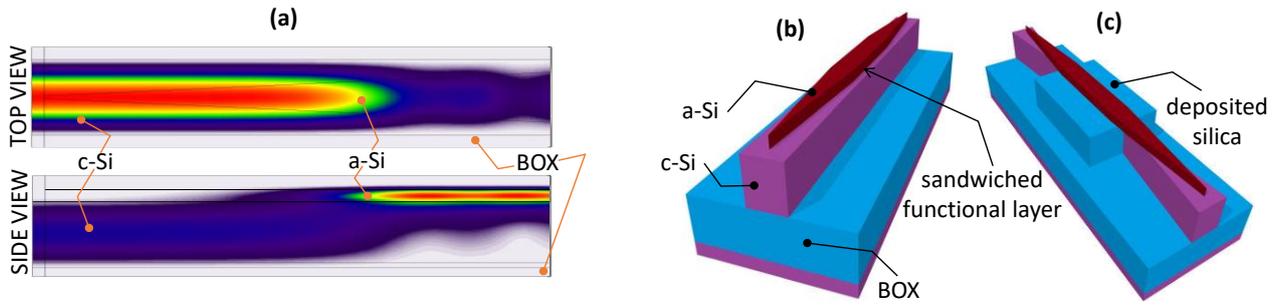

Figure 9. (a) 3D simulation with the eigenmode expansion method of the adiabatic power transfer from a 3 µm thick c-Si waveguide to a 400 nm thick and 200 µm long a-Si:H tapered waveguide fabricated on top; (b) 3D sketch of two escalators to couple light to the a-Si:H waveguide and then back to the 3 µm thick waveguide, showing where a functional layer can be sandwiched between the two silicon types in the region where the light is guided in a-Si:H; (c) a different type of escalator to couple light to submicron waveguides.

### 2.2 Active building blocks

We can divide the active elements in two main categories: electrical-to-optical converters (EOCs), which, in the thick-SOI platform, are basically all phase shifters (either thermo-optic or electro-optic) and optical-to-electrical converters (OECs), i.e., photodetectors.

*Phase shifters*

We implement thermo-optic phase shifters by implanting a thin silicon pedestal at the bottom of strip waveguides (Figure 10(a)). We usually cut away the remaining part of the pedestal to achieve lower power consumption, reaching about 25 mW per π-shift, with both rise time and decay time of about 15 µs (i.e., a speed of about 66 kHz). Very recently, we have also demonstrated ~2 mW per π-shift (not yet published) by fabricating the heaters on special cavity SOI wafers, which limits the heat flow through the substrate.

Placing the heaters in direct contact with the silicon layer ensures a significant reduction of thermal cross-talk[55] compared to heaters based on metal wires placed on top of the waveguide upper cladding. This is a major advantage for complex circuits requiring several thermo-optic phase shifters. By design, thermo-optic phase shifters come with no excess loss.

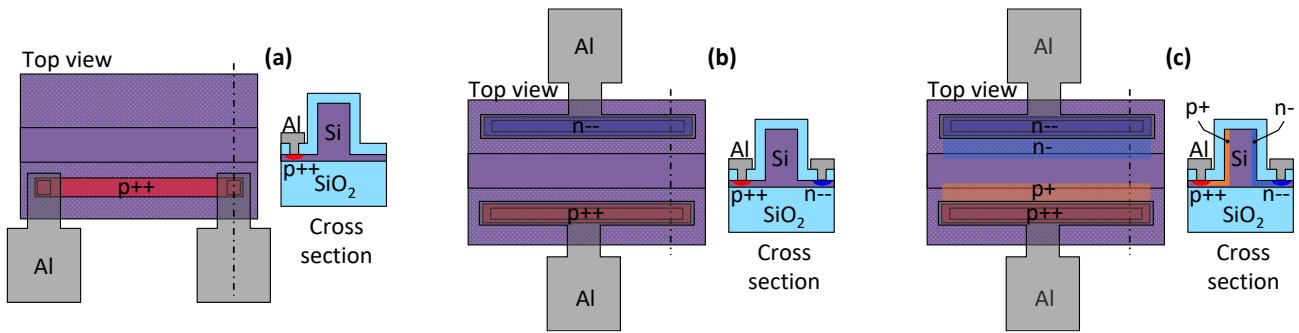

Figure 10. Top views and cross-sections of the three main types of phase shifters available on the platform: (a) thermo-optic (see also Figure 1); (b) electro-optic, based on plasma dispersion through carrier injection in a PIN junction; and (c) electro-optic, based on EFIPE with a high inverse bias voltage through a PIN junction.

When higher speed is needed, we can reach about 2 to 3 MHz using simple PIN junctions, with only one implantation level, as sketched in Figure 10(b). In this case the refractive index changes due to carrier injection inducing plasma dispersion[56]. The power consumption for a π-shift is lower than 5 mW. Nevertheless, plasma dispersion inherently adds amplitude modulation on top of the phase modulation, due to the Kramers-Kronig relations[29]. The loss associated with a π-shift is on the order of 1 dB to 2 dB. Indeed, when made sufficiently long, the same type of PIN junction is also used for variable optical attenuators[57].

To overcome these limitations, we are also developing phase modulators relying on the so called electric-field-induced Pockels effect (EFIPE, see Figure 10(c))[58], in close collaboration with the University of Tokyo. For these modulators, we expect significantly lower excess loss due to the high reverse bias (electric field of about 40 V/μm, as close as possible to the breakdown). In particular, we do not expect any major amplitude modulation associated with phase modulation. Furthermore, we aim to reach modulation speeds exceeding 1 GHz and possibly approaching 10 GHz. We also expect the power consumption to be in the microwatt range per π-shift, which is important for cryogenic applications. In fact, EFIPE works well also at cryogenic temperatures[59] because it is not affected by carrier freeze-out, unlike plasma dispersion[60].

With the goal of achieving extremely low power consumption in combination with modulation speeds exceeding 100 GHz, we are also developing plasmonic modulators in collaboration with ETH Zürich and the company Polariton Technologies[61]. Besides the conventional approach based on nonlinear polymers, we are also exploring the possible use of a-Si:H as nonlinear material based on EFIPE[62]. We point out that plasmonic modulators are also particularly suitable for cryogenic applications[63], since they do not rely on charge carriers and operate with ultra-low power dissipation[64]. The main limitation of plasmonic phase shifters is the high excess loss, typically exceeding 5 dB. However, as explained in §3, this can be acceptable in some applications.

To conclude this section, we point out that a key missing building block for quantum PICs (QPICs) in all platforms is a suitable phase shifter to simultaneously enable a small footprint, ultra-low power consumption, high speed, ultra-low optical loss, and cryogenic operation, or at least a subset of these properties, depending on the application. Recent results demonstrate that using microelectromechanical systems (MEMS) is a promising path for both submicron silicon[65] and silicon nitride[66,67] platforms. This approach can result in losses below 0.5 dB, speeds from a few MHz to beyond 100 MHz, and footprints ranging from about 100×100 μm$^2$ to 1×1 mm$^2$. One additional avenue, which we are presently exploring for faster and more compact phase shifters, is to place electro-optic metasurfaces[53,54] on top of URMs. Here the goal is to access the full nonlinear coefficient of electro-optic polymers, which is typically reduced by one order of magnitude in plasmonic slot waveguides[68].

*Detectors*

The platform includes monolithically integrated germanium (Ge) photodiodes (PD), with responsivity in the order of 1 A/W at 1550 nm wavelength, meaning 80% quantum efficiency. We have developed both high-speed PDs and monitor PDs. The high-speed PDs exceed 40 GHz speed when operated with 1 V reverse bias[10], with a dark current of about 4 μA, whereas monitor PDs are operated with lower bias voltage and have about 10 nA dark current with about 1 GHz speed. We are presently starting cryogenic characterisation of the PDs to determine the temperature dependence of dark current[69], responsivity, signal-to-noise ratio[70], speed, carrier freeze out, and wavelength range of detection. We are also developing avalanche photodetectors[71] (APDs) exploiting the avalanche effect in silicon[72] and also plan to operate them in Geiger mode to achieve single photon avalanche detectors (SPADs)[73].

Additionally, in collaboration with ETH Zürich, we are also developing high-speed plasmonic Ge detectors to exceed 500 GHz analog bandwidth[74,75]. Here the main driver is not the detection efficiency but high-speed operation at a few Kelvin, with the goal of developing suitable OECs to transfer a large amount of data to the cryostat. As explained in more detail in §4, the idea is to drive superconducting electronics (e.g., single flux quantum, SFQ) using optical fibres.

With cryogenic and quantum applications in mind, we are also developing guided-wave SNSPDs, with the final goal of coupling them through a light escalator (Figure 9(b)). In order to speed-up the detector development, we have started fabricating the devices sketched in Figure 11(c). We first oxidised a silicon wafer, deposited superconducting NbN on the thermal oxide, and patterned the nanowires using e-beam lithography, as shown in Figure 11(a). Next, we deposited a-Si:H and patterned the waveguides (Figure 11(b)) including inverse tapers to improve fibre coupling from the chip edge. The optical fibre is aligned using a nanopositioner in the cryostat. In parallel, we are also developing amorphous alternatives[76,77] to crystalline NbN, targeting improved fabrication yield of the detectors.

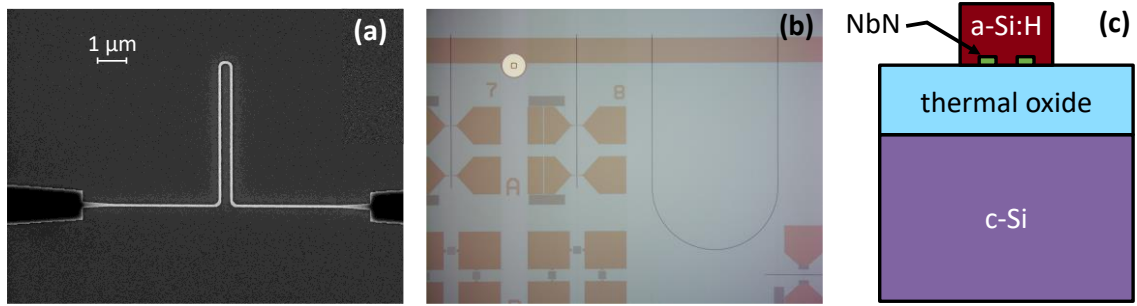

Figure 11. (a) SEM picture of a fabricated NbN SNSPD before a-Si:H deposition; (b) micrograph of a detail of a fabricated chip after etching the a-Si:H waveguides; (c) sketched cross-section of an a-Si:H waveguide with the NbN nanowire embedded (in green).

The SNSPD is the closest thing to an ideal single-photon detector demonstrated to date, with detection efficiencies exceeding 97% extending into the telecom wavelength range[78], speeds in the GHz range[79], jitter even lower than 3 ps[80], and dark counts under 0.1 Hz[81]. However, for some quantum realisations[82–84], and especially those based on Gaussian states, photon number resolution (PNR) is an important capability which is not easily addressed by SNSPDs. A possible solution is transition edge sensors (TESs)[85], but their speed is presently limited to about 1 MHz. Unfortunately, TESs require temperatures on the order of 100 mK, which cannot be achieved with closed-cycle table-top cryostats but require more complex and bulky sub-Kelvin coolers such as dilution refrigerators.

However, it should be noted here that solid-state cooling technology developed currently at VTT can provide a viable solution to integrate also compact sub-Kelvin refrigerators in the future[86,87]. These coolers are superconductor - silicon hybrid chips that can be directly connected to the PIC with 3D-integration schemes, such as flip-chip bonding. This would enable sub-Kelvin operation temperature for the circuits inside simple, cost-effective, and compact pulse-tube systems.

Integration of TESs on optical waveguides has already been demonstrated on other platforms[88,89]. Our integration approach is in line with what we explained above for SNSPDs and leverages the TES pixel[90] and SQUID-based readout technology[91] developed at VTT. From a broader perspective, the in-house integrated superconducting device technology based on Nb cross-junctions[92] can provide interesting opportunities for the quantum upgrade of the micron-scale platform (see also §4). For example, we presently use the technology for the TES readout circuits[89], SQUID magnetometers[93] and different Josephson parametric devices[94–96]. Furthermore, in collaboration with the Royal Institute of Technology KTH and with the company Single Quantum, we are also exploring the possible use of single SNSPDs as efficient PNR detectors[97–99].

## 2.3 Hybrid integration

Several different PIC technologies are available, including most mature submicron and micron-scale silicon and silicon nitride[20] platforms, and micron-scale indium phosphide[100] platforms, as well as the more recent lithium niobate on insulator platforms[101–104] and compound-on-insulator platforms[105]. Each material system comes with its strengths and weaknesses, and suitable combinations of complementary systems are thus often needed to achieve fully-integrated solutions. A quintessential case is the lack of monolithically integrated light sources in all platforms not based on compound semiconductors, where either heterogeneous integration[106] or hybrid integration[7,107] is needed to generate light on chip.

Our main focus at VTT is on hybrid integration based on high-precision flip-chip bonding at the wafer scale, which is suitable for mid-volume production in a CMOS fabrication facility like ours. Unlike with heterogeneous and monolithic integration, in the hybrid approach the silicon process and the III-V process (or the process of any other complementary material system) can happen in parallel in two different fabrication facilities, which comes with several advantages. These include shorter overall lead time, reduced process flow complexity, reduced constraints and trade-offs for the two material systems, and decoupled yield of the two processes, resulting in higher overall yield, i.e., higher cost efficiency. Another advantage of hybrid integration is that it is not bound to the constraints that may arise in monolithic integration regarding, e.g., CMOS compatibility and thermal budgets. Furthermore, hybrid integration can be made with commercially available dies (e.g., light sources or photodetectors) which can lead to even higher cost efficiency.

By using either vertical facets or URMs, we can easily integrate devices where light propagates respectively in-plane – like distributed Bragg reflector lasers, semiconductor optical amplifiers, or electro-absorption modulators – or out-of-plane – like vertical cavity emitting lasers or free-space photodetectors. In particular, the URM can be a key component for QPICs, as it is extremely low-loss, broadband, and polarisation-independent. For example, it can be used to efficiently couple light from deterministic single photon sources based on quantum dots in vertical cavities[108] or to couple single photons or Gaussian states to arrays of short SNSPDs (see Figure 12b). We stress here that, even though we have a clear path to monolithic integration of SNSPDs (see §2.2), based on the above considerations, hybrid integration will be the most efficient integration approach for large SNSPD arrays until we develop a SNSPD fabrication process with sufficiently high yield.

## 3. QKD RECEIVERS

A first example application that can be enabled by the thick-SOI platform is quantum key distribution (QKD) networks[109] with higher key rates and/or longer working distance. In fact, PIC solutions are in the product roadmap of major QKD players[110], because of their unmatched stability and scalability. Indeed, several examples of PIC-based implementations have been reported to date, covering different types of QKD schemes[111–119]. We have identified a clear path for how the platform could support the development and large-scale deployment of high-performance QKD systems for both discrete variable (DV) QKD and continuous variable (CV) QKD, as briefly presented in the following.

### 3.1 DV-QKD

DV-QKD systems are most suitable to cover long distances. The longest QKD link reported to date reached 830 km using a special configuration with a central node[120,121], whereas the longest point-to-point link exceeded 400 km[122] (corresponding to about 70 dB loss in ultra-low loss fibres). The best commercial systems are typically limited to the 100 km to 150 km range, mainly to ensure secure communication with key rates high enough to be useful. In fact, in the DV-QKD implementations most suitable for long distances, the key rate scales linearly with the link transmission probability $\eta$, which is the probability of a transmitted photon being detected at the receiver, accounting for all possible transmission and coupling losses as well as the limited detector efficiency[123]. In stark contrast to classical optical communication links, such key rate scaling implies a large mismatch between the transmission speed and the detection speed. In other words, the detector can be orders of magnitude slower than the modulator, which is a unique opportunity to combine the fastest ever achieved optical modulators with the most efficient single photon detectors demonstrated to date, namely plasmonic modulators and SNSPDs respectively. Transmitter speeds of present QKD systems are on the order of a few GHz, meaning that plasmonic phase and amplitude modulators could be used to boost the key rate by at least two orders of magnitude, while being still well matched by SNSPDs on the receiver side. In fact, present commercial SNSPDs can easily exceed 10 MHz count rates with more than 80% detection efficiency and will possibly exceed GHz count rates and 95% detection efficiency in the future. We stress that the high losses of plasmonic modulators, which are a strong limitation for classical optical communication, are not at all a problem for practical DV-QKD transmitters, which are based on strongly attenuated light sources. On the other hand, high losses are not tolerable for the receiver, so plasmonic modulators are not an option for protocols (like the standard BB84) where modulators are also needed to choose the measurement basis on the receiver side. Nevertheless, this is not a strong limitation, given that the most robust protocols for practical DV-QKD rely on passive receivers, requiring no modulators[122,124,125].

The combination of plasmonic modulators and SNSPDs becomes even more attractive when considering that some of the most promising DV-QKD protocols, including measurement-device-independent (MDI) QKD[126,127] and twin-field (TF) QKD[120,123], connect the users through a central unit (completely untrusted) where all the photon detections occur (Figure 12(a)). Large scale deployment of these systems can be achieved by providing all users with low-cost transmitters (achievable with plasmonic chips) while deploying a central detection unit – owned by the operator – to host a table-top closed-cycle cryostat where thousands of SNSPDs can be economically cooled down and operated in parallel. In this vision, the cryostat would be connected to tens to hundreds of fibres, and each fibre should carry tens to hundreds of wavelength division multiplexed (WDM) signals.

To this end, at VTT we are presently fabricating low loss AWGs to demultiplex the WDM signals coming from a single fibre and couple them to flip-chip-bonded arrays of SNSPDs designed and fabricated by Single Quantum to match our layout (see Figure 12b). Monolithic integration of AWGs and SNSPDs has been already demonstrated[128] but with high losses both for fibre coupling and demultiplexing. Furthermore, monolithic integration of large SNSPD arrays is still challenging, due to the relatively poor SNSPD fabrication yield. Hybrid integration of SNSPD chips (with only two detectors) has been recently demonstrated with submicron silicon waveguides[129] for time multiplexed MDI-QKD. The coupling losses demonstrated therein were very high, as grating couplers were used to couple both the optical fibre and the SNSPDs. We instead aim at a solution simultaneously ensuring high yield, broadband low-loss fibre coupling, low demultiplexing loss, and which can be even made polarization-insensitive with a suitable design of the SNSPD geometry[130,131]. The final goal is full monolithic integration of a DV-QKD receiver on the thick-SOI platform, providing much better and more stable control of relative phase and time jitter, therefore leading to higher fringe visibility.

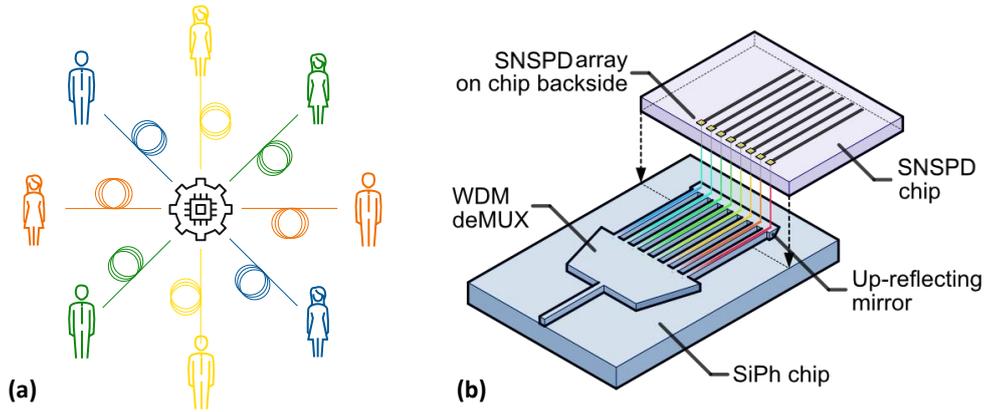

Figure 12. (a) Schematic representation of QKD implementations based on a central node for photon detection where all the users are equipped with suitable and low-cost transmitters; (b) 3D sketch of the solution we are developing with our partner Single Quantum to address arrays of SNSPDs with low loss and high fabrication yield.

## 3.2 CV-QKD

An alternative approach is CV-QKD, which relies on Gaussian states instead of single photons. The main advantage is that its implementation[41] requires only standard telecom components used for classical coherent optical communication, and, in particular, there is no need for single photon detectors. The main drawback is that secure implementations scale quadratically with the transmission probability $\eta$, which limits the operation range to about 50 km (or to be more rigorous, 10 dB loss, assuming standard 0.2 dB/km fibre loss). Furthermore, unlike DV-QKD, the receiver speed must match the transmitter speed. On the transmitter side, plasmonic modulators are again the perfect choice, given that their losses can be easily tolerated, and they can easily achieve both phase and amplitude ultrafast modulation simultaneously[64]. Ultrafast phase modulation is needed also on the receiver side but only on the local oscillator and not on the quantum states[41], meaning that some modulator losses are acceptable. Detection is typically done using shot-noise-limited balanced pulsed homodyne detectors[41] whose operation speed and stability can greatly benefit from PIC integration and dedicated electronics[132].

We therefore plan to exploit fast Ge PDs in combination with our in-house expertise in ultra-fast electronics[133,134] to develop balanced photodetectors with a speed above 50 GHz. We stress that, even though the speed of our present Ge PDs is limited to about 40 GHz, suitably designed Ge PDs with smaller volume have been recently demonstrated to reach up to 265 GHz[135]. In our vision, the CV-QKD receiver will be monolithically integrated on our thick SOI platform, to include the ultrafast plasmonic phase modulator and the balanced photodiode. Also in this case, the PIC will ensure much better and more stable control of relative phase and time jitter compared to realisations based on optical fibres, therefore leading to improved overall performance of the whole QKD system.

On the transmitter side, integration of the plasmonic devices on our platform would not be strictly necessary, but it could improve operation stability, for example through integrated Faraday mirrors (see §2.1) not available in any other PIC platform. Similar considerations apply to DV-QKD transmitters. Indeed, many practical implementations of both DV- and CV-QKD rely on Faraday mirrors[41,122,125].

The platform can also support quantum communication applications beyond QKD. A promising path is the development of acousto-optical devices to efficiently transduce superconducting qubits or spin qubits into optical qubits and vice versa. Such transducers would allow us to connect quantum processors via optical quantum states in optical fibres and create more powerful quantum computers based on multi-quantum-processor architectures, even using quantum computers located several kilometres away. With this application in mind, together with the University of Bristol, we are exploring the possible realisation of efficient piezoelectric microwave-to-optical transducers[136].

# 4. SCALING-UP CRYOGENIC QUANTUM COMPUTERS

A second example application is the use of optical fibres to transfer data to and from superconducting quantum computers, aiming to scale up the number of qubits and achieve useful universal quantum computing. We are currently in the 'Noisy Intermediate-Scale Quantum' (NISQ) era[137] – which means that significant applications are expected already in the short- and medium-term with a limited number of noisy qubits. However, it is generally agreed that universally useful quantum computers will require about one million qubits[138].

To date, the most advanced universal quantum computers are based on superconducting qubits and operated at temperatures below 50 mK, which is required to minimize thermal noise. A highly scalable approach based on silicon qubits is also quickly evolving[139–143] and requires low temperatures as well. In all cryogenic quantum processors, electrical transmission lines are used to carry the electrical signals driving and reading the qubits inside the cryostat. Even though this approach is feasible when dealing with a few hundreds of qubits, it becomes challenging for thousands of qubits and not viable anymore when approaching one million of qubits. In fact, electrical cables come with a detrimental trade-off between bandwidth and thermal conductivity. For these reasons, at VTT we are intensely developing the next generation of communication interfaces for cryogenic qubits, using optical fibres and suitable OECs and EOCs.

There are at least two significantly different research directions for the optical control of superconducting quantum technology: (*i*) A number of OECs generate the drive signals of a quantum computer at the cryogenic temperature. Here, the OEC must receive an optical signal from room temperature that is directly suitable for driving the qubits and their gates. (*ii*) A number of cryogenic OECs receive digital optical input signals and convert them into digital electrical signals driven into a superconducting single flux quantum (SFQ) device[144]. The SFQ is a superconducting processor for classical data that can also generate drive signals for the quantum computer, as shown in Figure 13. The packaging density (crucial for scaling up) of both approaches can be supported by integrated optical techniques, such as WDM for multiplexing multiple signals into the same optical fibre.

The control of qubits has already been demonstrated for the first approach[145]. This approach benefits from the possibility of using existing electrical qubit drive electronics whose signals are simply converted into an optical form with an EOC at room temperature. However, at the small signal levels required by quantum computers, OECs suffer from shot noise, which can be detrimental for driving their analogue signals into sensitive quantum computers. The second approach is significantly more tolerant to shot noise, since the OECs only need to generate digital signals for SFQ, which can generate quantized analogue signals based on digital input data.

Our vision follows this second approach, illustrated in Figure 13, where a large amount of data from a supercomputer is serialised by a suitable EOC and sent through an optical fibre to a cryogenic OEC to drive SFQ logic[144]. After inputting the data into the quantum processor, the SFQ co-processors use the calculation output to drive a suitable cryogenic EOC that, through another optical fibre, sends the results to a de-serialising OEC which communicates back to the supercomputer.

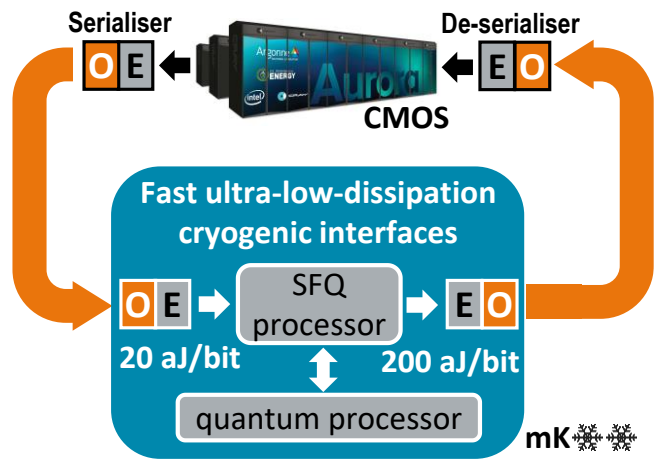

Figure 13. Schematic representation of our plans to use optical fibre links to interface cryogenic quantum computers with supercomputers.

The serialiser and de-serialiser are in general needed because the speed of SFQ logic is typically much higher than the speed of standard complementary metal oxide semiconductor (CMOS) electronics. SFQ is a promising choice due to its ultra-low energy dissipation, which is mandatory when working at the ultra-low temperatures required by superconducting quantum computers.

CMOS electronics can also be used in cryogenic environments, where low temperatures allow lower operating voltages and thus lower power consumption[146,147]. Cryo-CMOS uses traditional CMOS components that are tailored towards low temperature operation. However, with CMOS circuits, it is very hard to have sufficiently small dissipation. For example, the total power dissipation of only two spin qubit processor read-out and control circuitry operated at 3 K was 330 mW[148], which is already on high end of the tolerance level of the modern cryostats. Furthermore, reaching high enough clock rates ($> 1$ GHz) at low temperature is a significant challenge that, if not solved, implies higher qubit overhead. The strong points of the cryo-CMOS technology are the existing fabrication infrastructure and the advanced design tools and expertise.

More dramatic gains in energy efficiency are possible using single flux quantum (SFQ) technology. The latter, and its variants such as energy efficient SFQ, represent bits as short ($\approx 1$ ps) pulses produced by switching processes in superconducting tunnel junctions called Josephson junctions. The typical energy of these pulses is only 0.2 aJ, and the pulses can be processed at speeds exceeding 100 GHz[149]. In the past, their use has been limited by the requirement of cryogenic temperature operation and mediocre packing density of components. For qubit interfacing purposes neither of these issues is relevant. Superconducting electronics based on single flux quantum (SFQ) logic dissipates less than 1% of the energy dissipated by CMOS electronics[150] and enhanced variants (eSFQ or eRSFQ logic) can even dissipate less than 0.1% compared to the CMOS systems. SFQ controllers additionally enable vastly superior clock frequencies, in extreme cases even above 700 GHz[151]. With advanced thermal management, the SFQ controllers could be even directly integrated with the qubit processor altogether removing the need for very complex cabling solutions

As part of this vision, we are presently developing, together with our partners, several PIC solutions for different building blocks. For example, in Figure 14 we show a long-term vision of how to replace a prototype serialiser, presently based on optical fibres and discrete components, with a fully integrated PIC solution. A III-V reflective semiconductor optical amplifier (RSOA), including a saturable absorber (SA), is flip-chip bonded on the silicon chip where it is coupled to an integrated, compact, and low-loss external cavity to create an integrated mode-locked laser (IMLL). The generated wavelengths are then separated by a low-loss integrated demultiplexer. The signal in each waveguide is finally modulated independently through an array of amplitude modulators, each driven by relatively slow electrical signals (1 GHz to 2.5 GHz). The resulting signals are first delayed by multiples of a suitable delay unit, and finally recombined through a wavelength multiplexer. We are presently developing passive PICs combining the delay lines and the final multiplexer. We will test them as part of the free-space serialiser prototype we have already built.

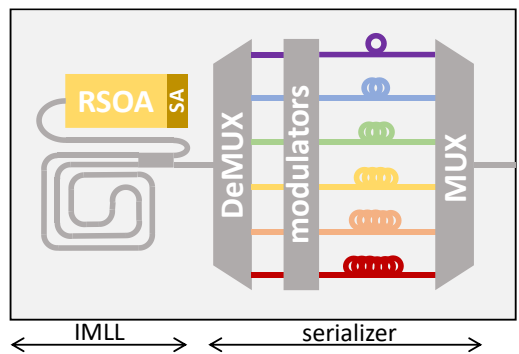

Figure 14. Long term vision of a PIC based serialiser, including an integrated mode-locked laser (IMLL) as multi-wavelength light source.

A second example is the cryogenic OEC that we are building using SNSPDs. In this particular application, we are more interested in their detection speed rather than extremely high detection efficiency, given that we can afford to use multiple photons per pulse. Together with our collaborators, we are trying to achieve the ultimate SNSPD speed. A simple approach is to make the nanowire as short as possible, but experimental results clearly show that latching[152–154] becomes an issue in doing so. Active electrical quenching has been also proposed, but without dramatic improvements[155]. In order for the OEC speed to approach the SFQ speed, we are also exploring different multiplexing approaches, addressing arrays of SNSPDs instead of single detectors. In this approach, we avoid using the optical serialiser at room temperature and replace it with an electrical serialiser inside the cryostat[156], resulting in major under-exploitation of the fibre bandwidth. The simplest brute force approach is space division multiplexing (SDM), i.e., coupling each SNSPD with a dedicated fibre. We are indeed developing 2D fibre arrays suitable for cryogenic illumination of detector arrays. A finer approach is to use WDM, the same way as in Figure 12(b), where a single fibre carries several wavelengths. Time division multiplexing (TDM) could be also an option, but it would require active control of a network of relatively fast switches.

The most demanding part of the vision in Figure 13 is by far the cryogenic EOC. In fact, the energy and the voltages available from the SFQ electronics are very low (attojoules per bit and microvolt respectively), and thus driving a fast optical modulator inside the cryostat is very challenging. Even though plasmonic modulators have been demonstrated to work with $< 1$ V and at attojoule the energy level[64], driving them with SFQ is still nontrivial and requires some major development, which we are presently tackling.

Together with our partners, we are presently developing the SFQ processors in parallel as well as PIC-enabled EOCs and OECs, and we plan to start testing combinations of these different building blocks in the next few years as proofs of concept of our vision. In the long run, this will support the development of the Finnish quantum computer, which has recently achieved the first milestone of five qubits[157] and is now targeting 50 qubit superconducting quantum computer by 2024. At the same time, the same optical interfacing technology will help scale up also the customized silicon qubit platform that we are co-developing[140,143].

## 5. OTHER INTERESTING APPLICATIONS AND CONCLUSIONS

To conclude, we briefly mention that the thick SOI technology can support many other quantum technology developments. For example, we have just started a project with KTH to integrate their thin lithium niobate waveguides[104] on our platform. We have also ongoing discussions with Tampere University on how to exploit the multimode behaviour and mode preservation capabilities of our PICs (see §2.1) to support spatial shaping of their qudits[34–36]. We have also identified turbulence mitigation[158] for satellite QKD as a promising application of our low-loss PICs with efficient phase shifters and integrated responsive detectors. We additionally have an ongoing collaboration with the Max Planck Institute for Quantum Optics to implant erbium in our silicon waveguides to achieve quantum emitters with narrow linewidth[159].

We have introduced the thick SOI platform with a special focus on the unique features that make it attractive for different quantum technologies, and we have also provided an overview of the ongoing developments to make it even more attractive in the near future. We presented two concrete cases elaborating in detail our vision of how PIC-based solutions will be able to support the large-scale deployment of high-performance QKD networks based on both DV-QKD and CV-QKD as well as the scaling-up of useful cryogenic quantum computers.

## BIOGRAPHIES

Matteo Cherchi received his Laurea in Physics from the University of Pavia (Italy) in 1997, the Certificate of Advanced Studies in Mathematics from the University of Cambridge (UK) in 1998, and his Dottorato di Ricerca in Electronic Engineering from the University of Palermo (Italy) in 2008. He has been working on photonic integration research since 1999, both in industry and academia. In 2002 he acted as visiting scientist at the Research Laboratory of Electronics at MIT, working closely with the groups of Prof. Hermann Haus and Prof. Lionel Kimerling. In 2011, he joined the microphotonics group of Prof. Michael R. Watts at MIT; he then moved to VTT's silicon photonics team led by Dr. Timo Aalto. He is presently Principal Scientist in the Quantum Photonics team at VTT.

Arijit Bera received his M. Sc. in Photonics (2013), as well as Ph.D. in Silicon Photonics (2018) from the Institute of Photonics, University of Eastern Finland, Joensuu, Finland. He joined VTT in 2018 as a research scientist, and started to work in the Silicon Photonics team since 2020. He has gained expertise in PIC design, and micro/nano fabrication of both active and passive photonic devices. He is current research interfaces the integrated photonics with the superconducting quantum technologies. Currently, he is with the Quantum Photonics team at VTT.

Antti Kemppinen received M. Sc. and D.Sc. degrees in Engineering Physics and Mathematics at Helsinki University of Technology in 2004 and 2009, respectively. He joined the Finnish national metrology institute in 2003, which was merged into VTT in 2015. Presently, he is a senior scientist in the Quantum Computing Hardware team of VTT where he coordinates European and national research projects. His research experience covers single-electron pumping, Josephson voltage standards, cryogenic thermometry, electrical refrigeration, superconducting nanowires, and cryogenic techniques including their optical integration.

Jaani Nissilä received M. Sc. and Dr. degrees in Engineering physics and mathematics at Helsinki University of Technology in 1994 and 2001, respectively. His studies focused on development of instrumentation and methods for defect studies in semiconductors using positron annihilation techniques. In 2002, he joined the Finnish national metrology institute to develop quantum standards for electrical metrology, especially Josephson voltage standards and their applications. Since 2015 he has been developing optical methods for driving superconducting electronics with fast optical pulses and ultrafast photodiodes.

Kirsi Tappura received the M.Sc. (Tech) (with distinction), Lic.Sc. (Tech), and D.Sc. (Tech.) degrees in technical physics from the Tampere University of Technology (TUT, currently Tampere University, TAU), Finland, in 1990, 1992, and 1993, respectively, while focusing on semiconductor physics and optoelectronics including quantum structures. After a period in industry among novel electronic display technologies, she joined VTT, where she has held various research and team leader positions and worked on a wide range of photonic/plasmonic, electronic and thermal devices based on semiconductor, superconductor and soft matter systems, with an emphasis on computational physics. Since 1999, she has also been Docent of Physics with TUT/TAU. She is currently Principal Scientist in the Quantum Sensors team at VTT.

Marco Caputo received his Bachelor's and Master's degree in Physics at the University of Salerno respectively in 2013 and 2015. In 2018 he obtained the Ph.D. in Mathematics, Physics and Applications at the University of Campania "Luigi Vanvitelli" in collaboration with the University of Salerno. During the time in academia, he studied the growth and characterization of superconducting films for spintronics and optoelectronics applications. In 2020 he joined the Quantum Hardware and Quantum Sensors teams at VTT as a Research Scientist focusing its activity on superconducting single-photon detectors. In 2022 he started working at Single Quantum as a Research Engineer.

Lauri Lehtimäki received his Master's degree in Theoretical physics from the University of Helsinki in 2018. After graduation, he worked in the field of photonic integration at VTT for three years. At VTT, he worked on the Silicon Photonics team, led by Dr. Timo Aalto, where his work concentrated on designing optical circuits and components on the silicon-on-insulator platform. Presently Lauri is working for a startup where he develops data analysis and signal processing methods.

Janne Lehtinen was research team leader of quantum sensors team at VTT during the manuscript writing and currently is chief science officer of Semiqon technologies Oy. He is expert in semiconductor and superconducting device fabrication and characterization. He has been in leading roles in European and national projects and published more than 30 peer-reviewed publications.

Joonas Govenius earned a B.A., summa cum laude, from Princeton University in 2010, a master's degree, with distinction, from ETH Zurich in 2012, and a doctoral degree from Aalto University in 2016, with a thesis on the design, fabrication and characterization of an ultrasensitive microwave bolometer. In 2018, he joined VTT and, since 2020, has led a research team of ca. 20 researchers focused on superconducting circuits, including qubits, sensors and amplifiers.

Tomi Hassinen received the M.Sc. and Ph.D. degrees in physics from the University of Helsinki in Helsinki, Finland in 2003 and 2018, respectively. He started working at VTT in 2006 and focused first on electronics production technologies. He is currently working as a Senior Scientist in Silicon Photonics team, doing research on component assembly and integration technologies. He has strong experience in component characterization and conducts the team's high speed component testing.

Mika Prunnila

Timo Aalto leads VTT's Silicon Photonics research team. His research focuses on the micron-scale silicon waveguides that are used to make ultra-compact and low-loss photonic integrated circuits for communication, imaging and sensing applications at infrared wavelengths. Aalto has contributed to approximately 100 scientific publications, reviewed several EU projects, journal articles and theses and coordinated large projects funded by EU, Business Finland, the European space agency and industry. Timo Aalto received his M.Sc. and D.Sc. (tech) degrees in optoelectronics technology from the Helsinki University of Technology in 1998 and 2004, respectively, and has worked in silicon photonics research at VTT since 1997.

## DISCLOSURES

The authors declare no conflicts of interest.

## ACKNOWLEDGMENTS

This work has been supported by the European Union's Horizon 2020 Research and Innovation Programme through projects aCryComm (Grant Agreement no. 899558), Quantum e-leaps (Grant Agreement no. 862660), OpenSuperQ (Grant Agreement no. 820363), EFINED (Grant Agreement no.766853), EMPIR SuperQuant (project no. 20FUN07). We acknowledge also support by the Academy of Finland Flagship Programme, Photonics Research and Innovation (PREIN), decision number 320168, and by the VTT internal quantum initiative "Quantum leap in quantum control". This work was performed as part of the Academy of Finland Centre of Excellence program (project 336817) and projects ETHEC (Grant Agreement no. 322580) and QuantLearn (Grant Agreement no. 350220). We also acknowledge financial support from Business Finland through projects QuTI (no. 40562/31/2020) and PICAP (no. 44065/31/2020).

We thank our colleagues Emma Mykkänen, Tapani Vehmas, Giovanni Delrosso, Katja Kohopää, Markku Kapulainen, Pekka Pursula, Mikko Kiviranta, and Himadri Majumdar as well as our collaborators Val Zwiller, Mario Castañeda, Stephan Steinhauer, Samuel Gyger, Eva De Leo, and Stefan Köpfli for their inputs and fruitful discussions.

## REFERENCES

[1] Acín, A., Bloch, I., Buhrman, H., Calarco, T., Eichler, C., Eisert, J., Esteve, D., Gisin, N., Glaser, S. J., Jelezko, F., Kuhr, S., Lewenstein, M., Riedel, M. F., Schmidt, P. O., Thew, R., Wallraff, A., Walmsley, I. and Wilhelm, F. K., "The quantum technologies roadmap: a European community view," New J. Phys. **20**(8), 080201 (2018).

[2] Wang, J., Sciarrino, F., Laing, A. and Thompson, M. G., "Integrated photonic quantum technologies," 5, Nat. Photonics **14**(5), 273–284 (2020).

[3] Elshaari, A. W., Pernice, W., Srinivasan, K., Benson, O. and Zwiller, V., "Hybrid integrated quantum photonic circuits," Nat. Photonics **14**(5), 285–298 (2020).

[4] Rodt, S. and Reitzenstein, S., "Integrated nanophotonics for the development of fully functional quantum circuits based on on-demand single-photon emitters," APL Photonics **6**(1), 010901 (2021).

[5] Pelucchi, E., Fagas, G., Aharonovich, I., Englund, D., Figueroa, E., Gong, Q., Hannes, H., Liu, J., Lu, C.-Y., Matsuda, N., Pan, J.-W., Schreck, F., Sciarrino, F., Silberhorn, C., Wang, J. and Jöns, K. D., "The potential and global outlook of integrated photonics for quantum technologies," 3, Nat. Rev. Phys. **4**(3), 194–208 (2022).

[6] Moody, G., Sorger, V. J., Blumenthal, D. J., Juodawlkis, P. W., Loh, W., Sorace-Agaskar, C., Jones, A. E., Balram, K. C., Matthews, J. C. F., Laing, A., Davanco, M., Chang, L., Bowers, J. E., Quack, N., Galland, C., Aharonovich, I., Wolff, M. A., Schuck, C., Sinclair, N., et al., "2022 Roadmap on integrated quantum photonics," J. Phys. Photonics **4**(1), 012501 (2022).

[7] Aalto, T., Cherchi, M., Harjanne, M., Bhat, S., Heimala, P., Sun, F., Kapulainen, M., Hassinen, T. and Vehmas, T., "Open-Access 3-μm SOI Waveguide Platform for Dense Photonic Integrated Circuits," IEEE J. Sel. Top. Quantum Electron. **25**(5), 1–9 (2019).

[8] Bera, A., Marin, Y., Harjanne, M., Cherchi, M. and Aalto, T., "Ultra-low loss waveguide platform in silicon photonics," 24 January 2022, San Francisco, CA, USA, 12006–4, SPIE.

[9] Marin, Y., Bera, A., Cherchi, M. and Aalto, T., "Ultra-High-Q Racetrack on Thick SOI Platform Through Hydrogen Annealing," ECOC 2022 48th Eur. Conf. Opt. Commun., We4E.3, Basel (2022).

[10] Vehmas, T., Kapulainen, M., Heimala, P., Delrosso, G., Sun, F., Gao, F. and Aalto, T., "Monolithic integration of up to 40 GHz Ge photodetectors in 3um SOI," Silicon Photonics XV **11285**, 112850V, International Society for Optics and Photonics (2020).

[11] Cherchi, M., Ylinen, S., Harjanne, M., Kapulainen, M. and Aalto, T., "Dramatic size reduction of waveguide bends on a micron-scale silicon photonic platform," Opt. Express **21**(15), 17814–17823 (2013).

[12] Zhang, B., Qubaisi, K. A., Cherchi, M., Harjanne, M., Ehrlichman, Y., Khilo, A. N. and Popović, M. A., "Compact multi-million Q resonators and 100 MHz passband filter bank in a thick-SOI photonics platform," Opt. Lett. **45**(11), 3005–3008 (2020).

[13] Shahwar, D., Cherchi, M., Harjanne, M., Kapulainen, M. and Aalto, T., "Polarization splitters for micron-scale silicon photonics," Silicon Photonics XVI **11691**, 1169104, SPIE (2021).
[14] Jalas, D., Hakemi, N., Cherchi, M., Harjanne, M., Petrov, A. and Eich, M., "Faraday rotation in silicon waveguides," 14th Int. Conf. Group IV Photonics, 141–142, IEEE, Berlin, Germany (2017).
[15] Gao, F., Ylinen, S., Kainlauri, M. and Kapulainen, M., "A Modified Bosch Process For Smooth Sidewall Etching," Proc. 22nd Micromechanics Microsyst. Technol. Eur. Workshop, 69–72, Vestfold University College, Toensberg, Norway (2011).
[16] Gao, F., Ylinen, S., Kainlauri, M. and Kapulainen, M., "Smooth silicon sidewall etching for waveguide structures using a modified Bosch process," J. MicroNanolithography MEMS MOEMS **13**(1), 013010–013010 (2014).
[17] Soref, R. A., Schmidtchen, J. and Petermann, K., "Large single-mode rib waveguides in GeSi-Si and Si-on-SiO2," IEEE J. Quantum Electron. **27**(8), 1971–1974 (1991).
[18] Dietrich, P.-I., Blaicher, M., Reuter, I., Billah, M., Hoose, T., Hofmann, A., Caer, C., Dangel, R., Offrein, B., Troppenz, U., Moehrle, M., Freude, W. and Koos, C., "In situ 3D nanoprinting of free-form coupling elements for hybrid photonic integration," Nat. Photonics **12**(4), 241–247 (2018).
[19] Aalto, T., "Broadband and polarization independent waveguide-fiber coupling," Silicon Photonics XVIII, 12426–12440, SPIE (2023).
[20] Rahim, A., Goyvaerts, J., Szelag, B., Fedeli, J.-M., Absil, P., Aalto, T., Harjanne, M., Littlejohns, C., Reed, G., Winzer, G., Lischke, S., Zimmermann, L., Knoll, D., Geuzebroek, D., Leinse, A., Geiselmann, M., Zervas, M., Jans, H., Stassen, A., et al., "Open-Access Silicon Photonics Platforms in Europe," IEEE J. Sel. Top. Quantum Electron. **25**(5), 1–18 (2019).
[21] Mashanovich, G. Z., Milosevic, M. M., Nedeljkovic, M., Owens, N., Xiong, B., Teo, E. J. and Hu, Y., "Low loss silicon waveguides for the mid-infrared," Opt. Express **19**(8), 7112–7119 (2011).
[22] Karioja, P., Alajoki, T., Cherchi, M., Ollila, J., Harjanne, M., Heinilehto, N., Suomalainen, S., Zia, N., Tuorila, H., Viheriälä, J., Guina, M., Buczyński, R., Kasztelanic, R., Salo, T., Virtanen, S., Kluczyński, P., Borgen, L., Ratajczyk, M. and Kalinowski, P., "Integrated multi-wavelength mid-IR light source for gas sensing," -Gener. Spectrosc. Technol. XI **10657**, 106570A, International Society for Optics and Photonics (2018).
[23] Lindner, C., Kunz, J., Herr, S. J., Kiessling, J., Wolf, S. and Kühnemann, F., "High-sensitivity quantum sensing with pump-enhanced spontaneous parametric down-conversion," arXiv:2208.07595 (2022).
[24] Rong, H., Liu, A., Nicolaescu, R., Paniccia, M., Cohen, O. and Hak, D., "Raman gain and nonlinear optical absorption measurements in a low-loss silicon waveguide," Appl. Phys. Lett. **85**(12), 2196–2198 (2004).
[25] Gil-Molina, A., Aldaya, I., Pita, J. L., Gabrielli, L. H., Fragnito, H. L. and Dainese, P., "Optical free-carrier generation in silicon nano-waveguides at 1550 nm," Appl. Phys. Lett. **112**(25), 251104 (2018).
[26] Morrison, B., Zhang, Y., Pagani, M., Eggleton, B. and Marpaung, D., "Four-wave mixing and nonlinear losses in thick silicon waveguides," Opt. Lett. **41**(11), 2418–2421 (2016).
[27] Pagani, M., Morrison, B., Zhang, Y., Casas-Bedoya, A., Aalto, T., Harjanne, M., Kapulainen, M., Eggleton, B. J. and Marpaung, D., "Low-error and broadband microwave frequency measurement in a silicon chip," Optica **2**(8), 751 (2015).
[28] Aalto, T., Harjanne, M., Ylinen, S., Kapulainen, M., Vehmas, T. and Cherchi, M., "Total internal reflection mirrors with ultra-low losses in 3 μm thick SOI waveguides," Proc SPIE **9367**, 93670B-93670B – 9 (2015).
[29] Saleh, B. E. A. and Teich, M. C., [Fundamentals of photonics], Wiley, New York (1991).
[30] Cherchi, M. and Aalto, T., "Bent optical waveguide," WO2014060648 A1 (2014).
[31] Jiang, X., Wu, H. and Dai, D., "Low-loss and low-crosstalk multimode waveguide bend on silicon," Opt. Express **26**(13), 17680–17689 (2018).
[32] Li, C., Liu, D. and Dai, D., "Multimode silicon photonics," Nanophotonics **8**(2), 227–247 (2019).
[33] Mohanty, A., Zhang, M., Dutt, A., Ramelow, S., Nussenzveig, P. and Lipson, M., "Quantum interference between transverse spatial waveguide modes," Nat. Commun. **8**(1), 14010 (2017).
[34] Brandt, F., Hiekkamäki, M., Bouchard, F., Huber, M. and Fickler, R., "High-dimensional quantum gates using full-field spatial modes of photons," Optica **7**(2), 98–107 (2020).
[35] Piccardo, M., Ginis, V., Forbes, A., Mahler, S., Friesem, A. A., Davidson, N., Ren, H., Dorrah, A. H., Capasso, F., Dullo, F. T., Ahluwalia, B. S., Ambrosio, A., Gigan, S., Treps, N., Hiekkamäki, M., Fickler, R., Kues, M., Moss, D., Morandotti, R., et al., "Roadmap on multimode light shaping," J. Opt. **24**(1), 013001 (2021).
[36] Hiekkamäki, M., Barros, R. F., Ornigotti, M. and Fickler, R., "Observation of the quantum Gouy phase," Nat. Photonics, 1–6 (2022).

[37] Wang, Y., Dai, D. and Dai, D., "Multimode silicon photonic waveguide corner-bend," Opt. Express **28**(7), 9062–9071 (2020).
[38] Aalto, T., "Devices and Methods for Polarization Splitting," WO2020089530A1 (2020).
[39] Harjanne, M., Aalto, T. and Cherchi, M., "Polarization Rotator," WO2020225479A1 (2020).
[40] Zbinden, H., Gautier, J. D., Gisin, N., Huttner, B., Muller, A. and Tittel, W., "Interferometry with Faraday mirrors for quantum cryptography," Electron. Lett. **33**(7), 586–588 (1997).
[41] Jouguet, P., Kunz-Jacques, S., Leverrier, A., Grangier, P. and Diamanti, E., "Experimental demonstration of long-distance continuous-variable quantum key distribution," Nat. Photonics **7**(5), 378–381 (2013).
[42] Martinelli, M., "A universal compensator for polarization changes induced by birefringence on a retracing beam," Opt. Commun. **72**(6), 341–344 (1989).
[43] Cherchi, M., Ylinen, S., Harjanne, M., Kapulainen, M. and Aalto, T., "MMI resonators based on metal mirrors and MMI mirrors: an experimental comparison," Opt. Express **23**(5), 5982–5993 (2015).
[44] Cherchi, M., Sun, F., Kapulainen, M., Harjanne, M. and Aalto, T., "Flat-top interleavers based on single MMIs," Silicon Photonics XV **11285**, 112850G, International Society for Optics and Photonics (2020).
[45] Cherchi, M., Sun, F., Kapulainen, M., Vehmas, T., Harjanne, M. and Aalto, T., "Fabrication tolerant flat-top interleavers," Proc SPIE **10108**, 101080V-101080V – 9 (2017).
[46] Bhat, S., Harjanne, M., Sun, F., Cherchi, M., Kapulainein, M., Hokkanen, A., Delrosso, G. and Aalto, T., "Low Loss Devices fabricated on the Open Access 3 μm SOI Waveguide Platform at VTT," presented at ECIO, 2019, Ghent.
[47] Bera, A., Cherchi, M., Tappura, K., Heimala, P. and Aalto, T., "Amorphous silicon waveguide escalator: monolithic integration of active components on 3-um SOI platform," Silicon Photonics XV **11285**, 1128507, International Society for Optics and Photonics (2020).
[48] Goldstein, J., Lin, H., Deckoff-Jones, S., Hempel, M., Lu, A.-Y., Richardson, K. A., Palacios, T., Kong, J., Hu, J. and Englund, D., "Waveguide-integrated mid-infrared photodetection using graphene on a scalable chalcogenide glass platform," 1, Nat. Commun. **13**(1), 3915 (2022).
[49] Giambra, M. A., Sorianello, V., Miseikis, V., Marconi, S., Montanaro, A., Galli, P., Pezzini, S., Coletti, C. and Romagnoli, M., "High-speed double layer graphene electro-absorption modulator on SOI waveguide," Opt. Express **27**(15), 20145–20155 (2019).
[50] Pernice, W. H. P., Schuck, C., Minaeva, O., Li, M., Goltsman, G. N., Sergienko, A. V. and Tang, H. X., "High-speed and high-efficiency travelling wave single-photon detectors embedded in nanophotonic circuits," Nat. Commun. **3**, 1325 (2012).
[51] Jiang, Z. H., Lin, L., Ma, D., Yun, S., Werner, D. H., Liu, Z. and Mayer, T. S., "Broadband and Wide Field-of-view Plasmonic Metasurface-enabled Waveplates," Sci. Rep. **4**(1), 7511 (2014).
[52] Khorasaninejad, M. and Capasso, F., "Metalenses: Versatile multifunctional photonic components," Science (2017).
[53] Benea-Chelmus, I.-C., Meretska, M. L., Elder, D. L., Tamagnone, M., Dalton, L. R. and Capasso, F., "Electro-optic spatial light modulator from an engineered organic layer," Nat. Commun. **12**(1), 5928 (2021).
[54] Benea-Chelmus, I.-C., Mason, S., Meretska, M. L., Elder, D. L., Kazakov, D., Shams-Ansari, A., Dalton, L. R. and Capasso, F., "Gigahertz free-space electro-optic modulators based on Mie resonances," ArXiv210803539 Phys. (2021).
[55] Sabouri, S., Mendoza Velasco, L. A., Catuneanu, M., Namdari, M. and Jamshidi, K., "Thermo Optical Phase Shifter With Low Thermal Crosstalk for SOI Strip Waveguide," IEEE Photonics J. **13**(2), 1–12 (2021).
[56] Soref, R. and Bennett, B., "Electrooptical effects in silicon," IEEE J. Quantum Electron. **23**(1), 123–129 (1987).
[57] Zheng, D. W., Smith, B. T. and Asghari, M., "Improved efficiency Si-photonic attenuator," Opt. Express **16**(21), 16754–16765 (2008).
[58] Timurdogan, E., Poulton, C. V., Byrd, M. J. and Watts, M. R., "Electric field-induced second-order nonlinear optical effects in silicon waveguides," Nat. Photonics **11**(3), 200–206 (2017).
[59] Chakraborty, U., Carolan, J., Clark, G., Bunandar, D., Gilbert, G., Notaros, J., Watts, M. R. and Englund, D. R., "Cryogenic operation of silicon photonic modulators based on the DC Kerr effect," Optica **7**(10), 1385 (2020).
[60] Gehl, M., Long, C., Trotter, D., Starbuck, A., Pomerene, A., Wright, J. B., Melgaard, S., Siirola, J., Lentine, A. L. and DeRose, C., "Operation of high-speed silicon photonic micro-disk modulators at cryogenic temperatures," Optica **4**(3), 374 (2017).

[61] Burla, M., Hoessbacher, C., Heni, W., Haffner, C., Fedoryshyn, Y., Werner, D., Watanabe, T., Massler, H., Elder, D. L., Dalton, L. R. and Leuthold, J., "500 GHz plasmonic Mach-Zehnder modulator enabling sub-THz microwave photonics," APL Photonics **4**(5), 056106 (2019).

[62] Cherchi, M., "Electro-Optic Plasmonic Devices," WO2021099686A1 (2021).

[63] Habegger, P., Horst, Y., Koepfli, S., Kohli, M., De Leo, E., Bisang, D., Destraz, M., Tedaldi, V., Meier, N., Del Medico, N., Wang, W., Hoessbacher, C., Baeuerle, B., Heni, W. and Leuthold, J., "Plasmonic 100-GHz Electro-Optic Modulators for Cryogenic Applications," ECOC 2022 48th Eur. Conf. Opt. Commun., Tu1G.1, Basel (2022).

[64] Heni, W., Fedoryshyn, Y., Baeuerle, B., Josten, A., Hoessbacher, C. B., Messner, A., Haffner, C., Watanabe, T., Salamin, Y., Koch, U., Elder, D. L., Dalton, L. R. and Leuthold, J., "Plasmonic IQ modulators with attojoule per bit electrical energy consumption," Nat. Commun. **10**(1), 1–8 (2019).

[65] Edinger, P., Takabayashi, A. Y., Errando-Herranz, C., Khan, U., Sattari, H., Verheyen, P., Bogaerts, W., Quack, N. and Gylfason, K. B., "Silicon photonic microelectromechanical phase shifters for scalable programmable photonics," Opt. Lett. **46**(22), 5671–5674 (2021).

[66] Grottke, T., Hartmann, W., Schuck, C. and Pernice, W. H. P., "Optoelectromechanical phase shifter with low insertion loss and a 13π tuning range," Opt. Express **29**(4), 5525–5537 (2021).

[67] Dong, M., Clark, G., Leenheer, A. J., Zimmermann, M., Dominguez, D., Menssen, A. J., Heim, D., Gilbert, G., Englund, D. and Eichenfield, M., "High-speed programmable photonic circuits in a cryogenically compatible, visible–near-infrared 200 mm CMOS architecture," Nat. Photonics, 1–7 (2021).

[68] Xu, H., Elder, D. L., Johnson, L. E., Heni, W., Coene, Y. de, Leo, E. D., Destraz, M., Meier, N., Ghinst, W. V., Hammond, S. R., Clays, K., Leuthold, J., Dalton, L. R. and Robinson, B. H., "Design and synthesis of chromophores with enhanced electro-optic activities in both bulk and plasmonic–organic hybrid devices," Mater. Horiz. (2021).

[69] Pizzone, A., Srinivasan, S. A., Verheyen, P., Lepage, G., Balakrishnan, S. and Van Campenhout, J., "Analysis of dark current in Ge-on-Si photodiodes at cryogenic temperatures," 2020 IEEE Photonics Conf. IPC, 1–2 (2020).

[70] Siontas, S., Li, D., Liu, P., Aujla, S., Zaslavsky, A. and Pacifici, D., "Low-Temperature Operation of High-Efficiency Germanium Quantum Dot Photodetectors in the Visible and Near Infrared," Phys. Status Solidi A **215**(3), 1700453 (2018).

[71] Zhang, Q., Fu, S., Man, J., Li, Z., Cherchi, M., Heimala, P., Harjanne, M., Fei, S., Hiltunen, M., Aalto, T. and Zeng, L., "Low-loss and polarization-insensitive photonic integrated circuit based on micron-scale SOI platform for high density TDM PONs," 2017 Opt. Fiber Commun. Conf. Exhib. OFC, 1–3 (2017).

[72] Martinez, N. J. D., Derose, C. T., Brock, R. W., Starbuck, A. L., Pomerene, A. T., Lentine, A. L., Trotter, D. C. and Davids, P. S., "High performance waveguide-coupled Ge-on-Si linear mode avalanche photodiodes," Opt. Express **24**(17), 19072–19081 (2016).

[73] Vines, P., Kuzmenko, K., Kirdoda, J., Dumas, D. C. S., Mirza, M. M., Millar, R. W., Paul, D. J. and Buller, G. S., "High performance planar germanium-on-silicon single-photon avalanche diode detectors," Nat. Commun. **10**(1), 1086 (2019).

[74] Salamin, Y., Ma, P., Baeuerle, B., Emboras, A., Fedoryshyn, Y., Heni, W., Cheng, B., Josten, A. and Leuthold, J., "100 GHz Plasmonic Photodetector," ACS Photonics **5**(8), 3291–3297 (2018).

[75] Koepfli, S., Eppenberger, M., Hossain, M. S.-B., Baumann, M., Doderer, M., Destraz, M., Habegger, P., De Leo, E., Heni, W., Hoessbacher, C., Baeuerle, B., Fedoryshyn, Y. and Leuthold, J., ">500 GHz Bandwidth Graphene photodetector Enabling Highest-Capacity Plasmonic-to-Plasmonic Links," ECOC 2022 48th Eur. Conf. Opt. Commun., Th3B.5, Basel (2022).

[76] Mykkänen, E., Bera, A., Lehtinen, J. S., Ronzani, A., Kohopää, K., Hönigl-Decrinis, T., Shaikhaidarov, R., de Graaf, S. E., Govenius, J. and Prunnila, M., "Enhancement of Superconductivity by Amorphizing Molybdenum Silicide Films Using a Focused Ion Beam," 5, Nanomaterials **10**(5), 950 (2020).

[77] Häußler, M., Mikhailov, M. Yu., Wolff, M. A. and Schuck, C., "Amorphous superconducting nanowire single-photon detectors integrated with nanophotonic waveguides," APL Photonics **5**(7), 076106 (2020).

[78] Chang, J., Los, J. W. N., Tenorio-Pearl, J. O., Noordzij, N., Gourgues, R., Guardiani, A., Zichi, J. R., Pereira, S. F., Urbach, H. P., Zwiller, V., Dorenbos, S. N. and Esmaeil Zadeh, I., "Detecting telecom single photons with 99.5−2.07+0.5% system detection efficiency and high time resolution," APL Photonics **6**(3), 036114 (2021).

[79] Beutel, F., Gehring, H., Wolff, M. A., Schuck, C. and Pernice, W., "Detector-integrated on-chip QKD receiver for GHz clock rates," Npj Quantum Inf. **7**(1), 40 (2021).

[80] Korzh, B., Zhao, Q.-Y., Allmaras, J. P., Frasca, S., Autry, T. M., Bersin, E. A., Beyer, A. D., Briggs, R. M., Bumble, B., Colangelo, M., Crouch, G. M., Dane, A. E., Gerrits, T., Lita, A. E., Marsili, F., Moody, G., Peña, C., Ramirez, E., Rezac, J. D., et al., "Demonstration of sub-3 ps temporal resolution with a superconducting nanowire single-photon detector," Nat. Photonics **14**(4), 250–255 (2020).

[81] Mueller, A. S., Mueller, A. S., Korzh, B., Korzh, B., Runyan, M., Wollman, E. E., Beyer, A. D., Allmaras, J. P., Allmaras, J. P., Velasco, A. E., Craiciu, I., Bumble, B., Briggs, R. M., Narvaez, L., Peña, C., Peña, C., Spiropulu, M. and Shaw, M. D., "Free-space coupled superconducting nanowire single-photon detector with low dark counts," Optica **8**(12), 1586–1587 (2021).

[82] Bourassa, J. E., Alexander, R. N., Vasmer, M., Patil, A., Tzitrin, I., Matsuura, T., Su, D., Baragiola, B. Q., Guha, S., Dauphinais, G., Sabapathy, K. K., Menicucci, N. C. and Dhand, I., "Blueprint for a Scalable Photonic Fault-Tolerant Quantum Computer," Quantum **5**, 392 (2021).

[83] Arrazola, J. M., Bergholm, V., Brádler, K., Bromley, T. R., Collins, M. J., Dhand, I., Fumagalli, A., Gerrits, T., Goussev, A., Helt, L. G., Hundal, J., Isacsson, T., Israel, R. B., Izaac, J., Jahangiri, S., Janik, R., Killoran, N., Kumar, S. P., Lavoie, J., et al., "Quantum circuits with many photons on a programmable nanophotonic chip," 7848, Nature **591**(7848), 54–60 (2021).

[84] Madsen, L. S., Laudenbach, F., Askarani, M. F., Rortais, F., Vincent, T., Bulmer, J. F. F., Miatto, F. M., Neuhaus, L., Helt, L. G., Collins, M. J., Lita, A. E., Gerrits, T., Nam, S. W., Vaidya, V. D., Menotti, M., Dhand, I., Vernon, Z., Quesada, N. and Lavoie, J., "Quantum computational advantage with a programmable photonic processor," 7912, Nature **606**(7912), 75–81 (2022).

[85] You, L., "Superconducting nanowire single-photon detectors for quantum information," Nanophotonics **9**(9), 2673–2692 (2020).

[86] Mykkänen, E., Lehtinen, J. S., Grönberg, L., Shchepetov, A., Timofeev, A. V., Gunnarsson, D., Kemppinen, A., Manninen, A. J. and Prunnila, M., "Thermionic junction devices utilizing phonon blocking," Sci. Adv. **6**(15), eaax9191 (2020).

[87] Kemppinen, A., Ronzani, A., Mykkänen, E., Hätinen, J., Lehtinen, J. S. and Prunnila, M., "Cascaded superconducting junction refrigerators: Optimization and performance limits," Appl. Phys. Lett. **119**(5), 052603 (2021).

[88] Höpker, J. P., Gerrits, T., Lita, A., Krapick, S., Herrmann, H., Ricken, R., Quiring, V., Mirin, R., Nam, S. W., Silberhorn, C. and Bartley, T. J., "Integrated transition edge sensors on titanium in-diffused lithium niobate waveguides," APL Photonics **4**(5), 056103 (2019).

[89] Calkins, B., Mennea, P. L., Lita, A. E., Metcalf, B. J., Kolthammer, W. S., Lamas-Linares, A., Spring, J. B., Humphreys, P. C., Mirin, R. P., Gates, J. C., Smith, P. G. R., Walmsley, I. A., Gerrits, T. and Nam, S. W., "High quantum-efficiency photon-number-resolving detector for photonic on-chip information processing," Opt. Express **21**(19), 22657 (2013).

[90] Palosaari, M. R. J., Gronberg, L., Kinnunen, K. M., Gunnarsson, D., Prunnila, M. and Maasilta, I. J., "Large 256-Pixel X-ray Transition-Edge Sensor Arrays With Mo/TiW/Cu Trilayers," IEEE Trans. Appl. Supercond. **25**(3), 1–4 (2015).

[91] Akamatsu, H., Vaccaro, D., Gottardi, L., van der Kuur, J., de Vries, C. P., Kiviranta, M., Ravensberg, K., D'Andrea, M., Taralli, E., de Wit, M., Bruijn, M. P., van der Hulst, P., den Hartog, R. H., van Leeuwen, B.-J., van der Linden, A. J., McCalden, A. J., Nagayoshi, K., Nieuwenhuizen, A. C. T., Ridder, M. L., et al., "Demonstration of MHz frequency domain multiplexing readout of 37 transition edge sensors for high-resolution x-ray imaging spectrometers," Appl. Phys. Lett. **119**(18), 182601 (2021).

[92] Grönberg, L., Kiviranta, M., Vesterinen, V., Lehtinen, J., Simbierowicz, S., Luomahaara, J., Prunnila, M. and Hassel, J., "Side-wall spacer passivated sub-$\mu$m Josephson junction fabrication process," Supercond. Sci. Technol. **30**(12), 125016 (2017).

[93] Luomahaara, J., Kiviranta, M., Gronberg, L., Zevenhoven, K. C. J. and Laine, P., "Unshielded SQUID Sensors for Ultra-Low-Field Magnetic Resonance Imaging," IEEE Trans. Appl. Supercond. **28**(4), 1–4 (2018).

[94] Simbierowicz, S., Vesterinen, V., Grönberg, L., Lehtinen, J., Prunnila, M. and Hassel, J., "A flux-driven Josephson parametric amplifier for sub-GHz frequencies fabricated with side-wall passivated spacer junction technology," Supercond. Sci. Technol. **31**(10), 105001 (2018).

[95] Simbierowicz, S., Vesterinen, V., Milem, J., Lintunen, A., Oksanen, M., Roschier, L., Grönberg, L., Hassel, J., Gunnarsson, D. and Lake, R. E., "Characterizing cryogenic amplifiers with a matched temperature-variable noise source," Rev. Sci. Instrum. **92**(3), 034708 (2021).

[96] Perelshtein, M., Petrovnin, K., Vesterinen, V., Raja, S. H., Lilja, I., Will, M., Savin, A., Simbierowicz, S., Jabdaraghi, R., Lehtinen, J., Grönberg, L., Hassel, J., Prunnila, M., Govenius, J., Paraoanu, S. and Hakonen, P., "Broadband continuous variable entanglement generation using Kerr-free Josephson metamaterial," ArXiv211106145 Quant-Ph (2021).

[97] Single Quantum., "Application Note: Photon Number Resolving Detectors" (2021).

[98] Cahall, C., Nicolich, K. L., Islam, N. T., Lafyatis, G. P., Miller, A. J., Gauthier, D. J. and Kim, J., "Multi-photon detection using a conventional superconducting nanowire single-photon detector," Optica **4**(12), 1534–1535 (2017).

[99] Cheng, R., Zhou, Y., Wang, S., Shen, M., Taher, T. and Tang, H. X., "A 100-pixel photon-number-resolving detector unveiling photon statistics," 1, Nat. Photonics **17**(1), 112–119 (2023).

[100] Hoefler, G. E., Zhou, Y., Anagnosti, M., Bhardwaj, A., Abolghasem, P., James, A., Luna, S., Debackere, P., Dentai, A., Vallaitis, T., Liu, P., Missey, M., Corzine, S., Evans, P., Lal, V., Ziari, M., Welch, D., Kish, F., Suelzer, J. S., et al., "Foundry Development of System-On-Chip InP-Based Photonic Integrated Circuits," IEEE J. Sel. Top. Quantum Electron. **25**(5), 1–17 (2019).

[101] Lin, J., Bo, F., Bo, F., Cheng, Y., Cheng, Y., Cheng, Y., Cheng, Y., Xu, J. and Xu, J., "Advances in on-chip photonic devices based on lithium niobate on insulator," Photonics Res. **8**(12), 1910–1936 (2020).

[102] Luke, K., Kharel, P., Reimer, C., He, L., Loncar, M., Loncar, M. and Zhang, M., "Wafer-scale low-loss lithium niobate photonic integrated circuits," Opt. Express **28**(17), 24452–24458 (2020).

[103] Obrzud, E., Denis, S., Sattari, H., Choong, G., Kundermann, S., Dubochet, O., Despont, M., Lecomte, S., Ghadimi, A. H. and Brasch, V., "Stable and compact RF-to-optical link using lithium niobate on insulator waveguides," APL Photonics **6**(12), 121303 (2021).

[104] Prencipe, A., Baghban, M. A. and Gallo, K., "Tunable Ultranarrowband Grating Filters in Thin-Film Lithium Niobate," ACS Photonics **8**(10), 2923–2930 (2021).

[105] Chang, L., Cole, G. D., Moody, G. and Bowers, J. E., "CSOI: Beyond Silicon-on-Insulator Photonics," Opt. Photonics News **33**(1), 24–32 (2022).

[106] Komljenovic, T., Huang, D., Pintus, P., Tran, M. A., Davenport, M. L. and Bowers, J. E., "Photonic Integrated Circuits Using Heterogeneous Integration on Silicon," Proc. IEEE **106**(12), 2246–2257 (2018).

[107] Kapulainen, M., Ylinen, S., Aalto, T., Harjanne, M., Solehmainen, K., Ollila, J. and Vilokkinen, V., "Hybrid integration of InP lasers with SOI waveguides using thermocompression bonding," 2008 5th IEEE Int. Conf. Group IV Photonics, 61–63 (2008).

[108] Somaschi, N., Giesz, V., Santis, L. D., Loredo, J. C., Almeida, M. P., Hornecker, G., Portalupi, S. L., Grange, T., Antón, C., Demory, J., Gómez, C., Sagnes, I., Lanzillotti-Kimura, N. D., Lemaître, A., Auffeves, A., White, A. G., Lanco, L. and Senellart, P., "Near-optimal single-photon sources in the solid state," 5, Nat. Photonics **10**(5), 340–345 (2016).

[109] Diamanti, E., Lo, H.-K., Qi, B. and Yuan, Z., "Practical challenges in quantum key distribution," Npj Quantum Inf. **2**, npjqi201625 (2016).

[110] Paraïso, T. K., De Marco, I., Roger, T., Marangon, D. G., Dynes, J. F., Lucamarini, M., Yuan, Z. and Shields, A. J., "A modulator-free quantum key distribution transmitter chip," Npj Quantum Inf. **5**(1), 1–6 (2019).

[111] Ma, C., Sacher, W. D., Tang, Z., Mikkelsen, J. C., Yang, Y., Xu, F., Thiessen, T., Lo, H.-K. and Poon, J. K. S., "Silicon photonic transmitter for polarization-encoded quantum key distribution," Optica **3**(11), 1274–1278 (2016).

[112] Sibson, P., Erven, C., Godfrey, M., Miki, S., Yamashita, T., Fujiwara, M., Sasaki, M., Terai, H., Tanner, M. G., Natarajan, C. M., Hadfield, R. H., O'Brien, J. L. and Thompson, M. G., "Chip-based quantum key distribution," 1, Nat. Commun. **8**(1), 1–6 (2017).

[113] Sibson, P., Kennard, J. E., Stanisic, S., Erven, C., O'Brien, J. L. and Thompson, M. G., "Integrated silicon photonics for high-speed quantum key distribution," Optica **4**(2), 172–177 (2017).

[114] Cai, H., Long, C. M., DeRose, C. T., Boynton, N., Urayama, J., Camacho, R., Pomerene, A., Starbuck, A. L., Trotter, D. C., Davids, P. S. and Lentine, A. L., "Silicon photonic transceiver circuit for high-speed polarization-based discrete variable quantum key distribution," Opt. Express **25**(11), 12282–12294 (2017).

[115] Bunandar, D., Lentine, A., Lee, C., Cai, H., Long, C. M., Boynton, N., Martinez, N., DeRose, C., Chen, C., Grein, M., Trotter, D., Starbuck, A., Pomerene, A., Hamilton, S., Wong, F. N. C., Camacho, R., Davids, P., Urayama, J. and Englund, D., "Metropolitan Quantum Key Distribution with Silicon Photonics," Phys. Rev. X **8**(2), 021009 (2018).

[116] Zhang, G., Haw, J. Y., Cai, H., Xu, F., Assad, S. M., Fitzsimons, J. F., Zhou, X., Zhang, Y., Yu, S., Wu, J., Ser, W., Kwek, L. C. and Liu, A. Q., "An integrated silicon photonic chip platform for continuous-variable quantum key distribution," 12, Nat. Photonics **13**(12), 839–842 (2019).

[117] Avesani, M., Calderaro, L., Schiavon, M., Stanco, A., Agnesi, C., Santamato, A., Zahidy, M., Scriminich, A., Foletto, G., Contestabile, G., Chiesa, M., Rotta, D., Artiglia, M., Montanaro, A., Romagnoli, M., Sorianello, V., Vedovato, F., Vallone, G. and Villoresi, P., "Full daylight quantum-key-distribution at 1550 nm enabled by integrated silicon photonics," 1, Npj Quantum Inf. **7**(1), 1–8 (2021).

[118] Paraïso, T. K., Roger, T., Marangon, D. G., De Marco, I., Sanzaro, M., Woodward, R. I., Dynes, J. F., Yuan, Z. and Shields, A. J., "A photonic integrated quantum secure communication system," 11, Nat. Photonics **15**(11), 850–856 (2021).

[119] Cao, L., Luo, W., Wang, Y. X., Zou, J., Yan, R. D., Cai, H., Zhang, Y., Hu, X. L., Jiang, C., Fan, W. J., Zhou, X. Q., Dong, B., Luo, X. S., Lo, G. Q., Wang, Y. X., Xu, Z. W., Sun, S. H., Wang, X. B., Hao, Y. L., et al., "Chip-Based Measurement-Device-Independent Quantum Key Distribution Using Integrated Silicon Photonic Systems," Phys. Rev. Appl. **14**(1), 011001 (2020).

[120] Pittaluga, M., Minder, M., Lucamarini, M., Sanzaro, M., Woodward, R. I., Li, M.-J., Yuan, Z. and Shields, A. J., "600-km repeater-like quantum communications with dual-band stabilization," Nat. Photonics **15**(7), 530–535 (2021).

[121] Wang, S., Yin, Z.-Q., He, D.-Y., Chen, W., Wang, R.-Q., Ye, P., Zhou, Y., Fan-Yuan, G.-J., Wang, F.-X., Chen, W., Zhu, Y.-G., Morozov, P. V., Divochiy, A. V., Zhou, Z., Guo, G.-C. and Han, Z.-F., "Twin-field quantum key distribution over 830-km fibre," 2, Nat. Photonics **16**(2), 154–161 (2022).

[122] Boaron, A., Boso, G., Rusca, D., Vulliez, C., Autebert, C., Caloz, M., Perrenoud, M., Gras, G., Bussières, F., Li, M.-J., Nolan, D., Martin, A. and Zbinden, H., "Secure Quantum Key Distribution over 421 km of Optical Fiber," Phys. Rev. Lett. **121**(19), 190502 (2018).

[123] Lucamarini, M., Yuan, Z. L., Dynes, J. F. and Shields, A. J., "Overcoming the rate–distance limit of quantum key distribution without quantum repeaters," Nature **557**(7705), 400–403 (2018).

[124] Stucki, D., Brunner, N., Gisin, N., Scarani, V. and Zbinden, H., "Fast and simple one-way quantum key distribution," Appl. Phys. Lett. **87**(19), 194108 (2005).

[125] Boaron, A., Korzh, B., Houlmann, R., Boso, G., Rusca, D., Gray, S., Li, M.-J., Nolan, D., Martin, A. and Zbinden, H., "Simple 2.5 GHz time-bin quantum key distribution," Appl. Phys. Lett. **112**(17), 171108 (2018).

[126] Comandar, L. C., Lucamarini, M., Fröhlich, B., Dynes, J. F., Sharpe, A. W., Tam, S. W.-B., Yuan, Z. L., Penty, R. V. and Shields, A. J., "Quantum key distribution without detector vulnerabilities using optically seeded lasers," Nat. Photonics **10**(5), 312–315 (2016).

[127] Yin, H.-L., Chen, T.-Y., Yu, Z.-W., Liu, H., You, L.-X., Zhou, Y.-H., Chen, S.-J., Mao, Y., Huang, M.-Q., Zhang, W.-J., Chen, H., Li, M. J., Nolan, D., Zhou, F., Jiang, X., Wang, Z., Zhang, Q., Wang, X.-B. and Pan, J.-W., "Measurement-Device-Independent Quantum Key Distribution Over a 404 km Optical Fiber," Phys. Rev. Lett. **117**(19), 190501 (2016).

[128] Kahl, O., Ferrari, S., Kovalyuk, V., Vetter, A., Lewes-Malandrakis, G., Nebel, C., Korneev, A., Goltsman, G. and Pernice, W., "Spectrally multiplexed single-photon detection with hybrid superconducting nanophotonic circuits," Optica **4**(5), 557–562 (2017).

[129] Zheng, X., Zhang, P., Ge, R., Lu, L., He, G., Chen, Q., Qu, F., Zhang, L., Cai, X., Lu, Y., Zhu, S. N., Wu, P. and Ma, X., "Heterogeneously integrated, superconducting silicon-photonic platform for measurement-device-independent quantum key distribution," Adv. Photonics **3**(5), 055002 (2021).

[130] Chi, X., Zou, K., Gu, C., Zichi, J., Cheng, Y., Hu, N., Lan, X., Chen, S., Lin, Z., Zwiller, V. and Hu, X., "Fractal superconducting nanowire single-photon detectors with reduced polarization sensitivity," Opt. Lett. **43**(20), 5017–5020 (2018).

[131] Reddy, D. V., Otrooshi, N., Nam, S. W., Mirin, R. P. and Verma, V. B., "Broadband polarization insensitivity and high detection efficiency in high-fill-factor superconducting microwire single-photon detectors," APL Photonics **7**(5), 051302 (2022).

[132] Bruynsteen, C., Vanhoecke, M., Bauwelinck, J. and Yin, X., "Integrated balanced homodyne photonic–electronic detector for beyond 20 GHz shot-noise-limited measurements," Optica **8**(9), 1146–1152 (2021).

[133] Forsten, H., Saijets, J. H., Kantanen, M., Varonen, M., Kaynak, M. and Piironen, P., "Millimeter-Wave Amplifier-Based Noise Sources in SiGe BiCMOS Technology," IEEE Trans. Microw. Theory Tech. **69**(11), 4689–4696 (2021).

[134] Varonen, M., Sheikhipoor, N., Gabritchidze, B., Cleary, K., Forstén, H., Rücker, H. and Kaynak, M., "Cryogenic W-Band SiGe BiCMOS Low-Noise Amplifier," 2020 IEEEMTT- Int. Microw. Symp. IMS, 185–188 (2020).

[135] Lischke, S., Peczek, A., Morgan, J. S., Sun, K., Steckler, D., Yamamoto, Y., Korndörfer, F., Mai, C., Marschmeyer, S., Fraschke, M., Krüger, A., Beling, A. and Zimmermann, L., "Ultra-fast germanium photodiode with 3-dB bandwidth of 265 GHz," Nat. Photonics **15**(12), 925–931 (2021).

[136] Balram, K. C. and Srinivasan, K., "Piezoelectric Optomechanical Approaches for Efficient Quantum Microwave-to-Optical Signal Transduction: The Need for Co-Design," Adv. Quantum Technol. **5**(3), 2100095 (2022).

[137] Preskill, J., "Quantum Computing in the NISQ era and beyond," Quantum **2**, 79 (2018).

[138] "A Preview of Bristlecone, Google's New Quantum Processor.", Google AI Blog.

[139] Gonzalez-Zalba, M. F., de Franceschi, S., Charbon, E., Meunier, T., Vinet, M. and Dzurak, A. S., "Scaling silicon-based quantum computing using CMOS technology," Nat. Electron. **4**(12), 872–884 (2021).

[140] Duan, J., Lehtinen, J. S., Fogarty, M. A., Schaal, S., Lam, M. M. L., Ronzani, A., Shchepetov, A., Koppinen, P., Prunnila, M., Gonzalez-Zalba, F. and Morton, J. J. L., "Dispersive readout of reconfigurable ambipolar quantum dots in a silicon-on-insulator nanowire," Appl. Phys. Lett. **118**(16), 164002 (2021).

[141] Neyens, S., "High volume cryogenic measurement of silicon spin qubit devices from a 300 mm process line," 2022 Silicon Quantum Electron. Workshop, Orford (2022).

[142] Philips, S. G. J., Mądzik, M. T., Amitonov, S. V., de Snoo, S. L., Russ, M., Kalhor, N., Volk, C., Lawrie, W. I. L., Brousse, D., Tryputen, L., Wuetz, B. P., Sammak, A., Veldhorst, M., Scappucci, G. and Vandersypen, L. M. K., "Universal control of a six-qubit quantum processor in silicon," 7929, Nature **609**(7929), 919–924 (2022).

[143] Bohuslavskyi, H., Ronzani, A., Hätinen, J., Rantala, A., Shchepetov, A., Koppinen, P., Prunnila, M. and Lehtinen, J. S., "Scalable on-chip multiplexing of low-noise silicon electron and hole quantum dots," arXiv:2208.12131 (2022).

[144] Holmes, D. S., Ripple, A. L. and Manheimer, M. A., "Energy-Efficient Superconducting Computing—Power Budgets and Requirements," IEEE Trans. Appl. Supercond. **23**(3), 1701610–1701610 (2013).

[145] Lecocq, F., Quinlan, F., Cicak, K., Aumentado, J., Diddams, S. A. and Teufel, J. D., "Control and readout of a superconducting qubit using a photonic link," Nature **591**(7851), 575–579 (2021).

[146] Patra, B., Incandela, R. M., Dijk, J. P. G. van, Homulle, H. A. R., Song, L., Shahmohammadi, M., Staszewski, R. B., Vladimirescu, A., Babaie, M., Sebastiano, F. and Charbon, E., "Cryo-CMOS Circuits and Systems for Quantum Computing Applications," IEEE J. Solid-State Circuits **53**(1), 309–321 (2018).

[147] Bardin, J. C., Jeffrey, E., Lucero, E., Huang, T., Naaman, O., Barends, R., White, T., Giustina, M., Sank, D., Roushan, P., Arya, K., Chiaro, B., Kelly, J., Chen, J., Burkett, B., Chen, Y., Dunsworth, A., Fowler, A., Foxen, B., et al., "A 28nm Bulk-CMOS 4-to-8GHz ¡2mW Cryogenic Pulse Modulator for Scalable Quantum Computing," 2019 IEEE Int. Solid- State Circuits Conf. - ISSCC, 456–458 (2019).

[148] Xue, X., Russ, M., Samkharadze, N., Undseth, B., Sammak, A., Scappucci, G. and Vandersypen, L. M. K., "Quantum logic with spin qubits crossing the surface code threshold," 7893, Nature **601**(7893), 343–347 (2022).

[149] Likharev, K. K. and Semenov, V. K., "RSFQ logic/memory family: a new Josephson-junction technology for sub-terahertz-clock-frequency digital systems," Appl. Supercond. IEEE Trans. On **1**(1), 3–28 (1991).

[150] Mukhanov, O. A., "Energy-Efficient Single Flux Quantum Technology," IEEE Trans. Appl. Supercond. **21**(3), 760–769 (2011).

[151] Chen, W., Rylyakov, A. V., Patel, V., Lukens, J. E. and Likharev, K. K., "Rapid single flux quantum T-flip flop operating up to 770 GHz," IEEE Trans. Appl. Supercond. **9**(2), 3212–3215 (1999).

[152] Lv, C., Zhang, W., You, L., Hu, P., Wang, H., Li, H., Zhang, C., Huang, J., Wang, Y., Yang, X., Wang, Z. and Xie, X., "Improving maximum count rate of superconducting nanowire single-photon detector with small active area using series attenuator," AIP Adv. **8**(10), 105018 (2018).

[153] Annunziata, A. J., Quaranta, O., Santavicca, D. F., Casaburi, A., Frunzio, L., Ejrnaes, M., Rooks, M. J., Cristiano, R., Pagano, S., Frydman, A. and Prober, D. E., "Reset dynamics and latching in niobium superconducting nanowire single-photon detectors," J. Appl. Phys. **108**(8), 084507 (2010).

[154] Münzberg, J., Vetter, A., Beutel, F., Hartmann, W., Ferrari, S., Pernice, W. H. P. and Rockstuhl, C., "Superconducting nanowire single-photon detector implemented in a 2D photonic crystal cavity," Optica **5**(5), 658–665 (2018).

[155] Ravindran, P., Cheng, R., Tang, H., Bardin, J. C. and Bardin, J. C., "Active quenching of superconducting nanowire single photon detectors," Opt. Express **28**(3), 4099–4114 (2020).

[156] Miki, S., Miki, S., Miyajima, S., China, F., Yabuno, M. and Terai, H., "Photon detection at 1 ns time intervals using 16-element SNSPD array with SFQ multiplexer," Opt. Lett. **46**(24), 6015–6018 (2021).

[157] "Finland's first 5-qubit quantum computer is now operational.", <https://www.vttresearch.com/en/news-and-ideas/finlands-first-5-qubit-quantum-computer-now-operational> (3 January 2022 ).
[158] Billault, V., Bourderionnet, J., Mazellier, J. P., Leviandier, L., Feneyrou, P., Maho, A., Sotom, M., Normandin, X., Lonjaret, H. and Brignon, A., "Free space optical communication receiver based on a spatial demultiplexer and a photonic integrated coherent combining circuit," Opt. Express **29**(21), 33134–33143 (2021).
[159] Gritsch, A., Weiss, L., Früh, J., Rinner, S. and Reiserer, A., "Narrow optical transitions in erbium-implanted silicon waveguides," arXiv:2108.05120 (2021).